\documentclass{article}
\usepackage[final]{graphics}
\usepackage{amsfonts}

\font\tenmsa=msam10
\font\sevenmsa=msam7
\font\fivemsa=msam5
\font\tenmsb=msbm10
\font\sevenmsb=msbm7
\font\fivemsb=msbm5
\newfam\msafam
\newfam\msbfam
\textfont\msafam=\tenmsa  \scriptfont\msafam=\sevenmsa
\scriptscriptfont\msafam=\fivemsa
\textfont\msbfam=\tenmsb  \scriptfont\msbfam=\sevenmsb
\scriptscriptfont\msbfam=\fivemsb

\global\mathchardef\lesssim "142E

\newcommand{\slL}{\raise.15ex\hbox{$/$}\kern-.53em\hbox{$L$}}
\newcommand{\slP}{\raise.15ex\hbox{$/$}\kern-.53em\hbox{$P$}}
\newcommand{\slp}{\raise.1ex\hbox{$/$}\kern-.63em\hbox{$p$}}
\newcommand{\slk}{\raise.1ex\hbox{$/$}\kern-.63em\hbox{$k$}}
\newcommand{\slq}{\raise.1ex\hbox{$/$}\kern-.63em\hbox{$q$}}
\newcommand{\slv}{\raise.1ex\hbox{$/$}\kern-.63em\hbox{$v$}}
\newcommand{\slR}{\raise.15ex\hbox{$/$}\kern-.53em\hbox{$R$}}
\newcommand{\slQ}{\raise.15ex\hbox{$/$}\kern-.53em\hbox{$Q$}}
\newcommand{\slK}{\raise.15ex\hbox{$/$}\kern-.53em\hbox{$K$}}
\newcommand{\slSigma}{\raise.15ex\hbox{$/$}\kern-.53em\hbox{$\Sigma$}}
\newcommand{\slcalP}{\raise.15ex\hbox{$/$}\kern-.63em\hbox{$\cal P$}}
\newcommand{\slA}{\raise.15ex\hbox{$/$}\kern-.73em\hbox{$A$}}
\newcommand{\slbfA}{\raise.15ex\hbox{$/$}\kern-.73em\hbox{${\imb A}$}}
\newcommand{\slpartial}{\raise.15ex\hbox{$/$}\kern-.53em\hbox{$\partial$}}

\newcommand{\be}{\begin{equation}}
\newcommand{\ee}{\end{equation}}     
\newcommand{\bea}{\begin{eqnarray}}
\newcommand{\ena}{\end{eqnarray}}

\def\build#1\over#2{\mathrel{\mathop{\kern 0pt#1}\limits_{#2}}}

\font\tenimbf=cmmib10 at 10pt
\font\sevenimbf=cmmib10 at 7pt
\font\fiveimbf=cmmib10 at 5pt
\newfam\imbf
\textfont\imbf=\tenimbf
\scriptfont\imbf=\sevenimbf
\scriptscriptfont\imbf=\fiveimbf
\def\imb{\fam\imbf\tenimbf}

\begin{document}
\title{\bf{Coulomb corrections to $e^+e^-$ production\\ 
in ultra-relativistic nuclear collisions}}
\author{A.J.~Baltz, F.~Gelis, L.~McLerran, A.~Peshier\\
Brookhaven National Laboratory,\\
Physics Department, Nuclear Theory,\\
Upton, NY-11973, USA}
\maketitle

\begin{abstract}
  The purpose of this paper is to explain the discrepancies existing
  in the literature relative to $e^+e^-$ pair production in peripheral
  heavy ion collisions at ultra-relativistic energies. A controversial
  issue is the possible cancellation of Coulomb corrections to the
  Born term in the pair production cross-section.  Such a cancellation
  has been observed in a recent approach based on finding retarded
  solutions of the Dirac equation, but does not seem to hold in a
  perturbative approach. We show in this paper that the two approaches
  are in fact calculating different observables: the perturbative
  approach gives the exclusive cross-section of single pair
  production, while the other method gives the inclusive
  cross-section.
  
  We have also performed a thorough study of the electron propagator
  in the non-static background field of the two nuclei, the conclusion
  of which is that the retarded propagator is in the
  ultra-relativistic limit a much simpler object than the Feynman
  propagator, and can be calculated exactly.
\end{abstract}
\vskip 4mm 
\centerline{PACS codes: 25.75.-q, 11.80.-m, 34.90.+q, 11.55.-m\hfill BNL-NT-01/1}

\section{Introduction}
\begin{figure}[htbp]
\centerline{\resizebox*{1.7cm}{!}{\includegraphics{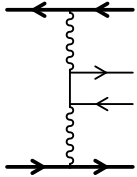}}}
\caption{\sl Born contribution to the pair production amplitude.}
\label{fig:born}
\end{figure}

In the past three years, the problem of $e^+e^-$ pair production
induced by the collision of two nuclei has attracted a lot of
interest, due partly to a series of papers showing that the Dirac
equation of an electron can be solved exactly in the electromagnetic
background field created by the two nuclei, in the limit where the two
nuclei are ultra-relativistic
\cite{SegevW1,SegevW2,BaltzM1,EichmRSG1}. This solution was then used
to derive an expression for the pair production cross-section. An
unexpected consequence of this formula was that it lead to a pair
production cross-section equal to its Born value (given by the diagram
in figure \ref{fig:born}).  In other words, this result seems to
indicate that all the corrections due to the multiple exchange of
photons (called ``Coulomb corrections'' in the following) cancel
exactly in the total cross-section.

This observation then prompted another series of papers on the same
topic, basically saying that this property cannot be true. One of them
\cite{IvanoM1} re-derived with new methods the classical result of
Davies, Bethe and Maximon \cite{BetheM1,BetheM2,DavieBM1} about the
fragmentation of a photon into a pair, induced by a electromagnetic
background field.  Then, this result was applied to pair production in
nuclear collisions \cite{IvanoSS1,IvanoSS2}. In this perturbative
calculation, the Coulomb corrections do not cancel and the resummed
cross-section is about $25\%$ smaller than the Born cross-section at
RHIC energy.  Others claimed the discrepancy to be due to a breakdown
of crossing symmetry \cite{EichmRG1}. Another paper \cite{LeeM1}
attributed the contradiction to a mishandling of ill-defined
integrals, and claimed that the result of Ivanov et al.  can be
recovered from the solution of the Dirac equation if a proper
regularization is used. But those papers apparently focused on
technical details, without first asking if an agreement between
\cite{SegevW1,SegevW2,BaltzM1,EichmRSG1} and \cite{IvanoSS1,IvanoSS2}
should be expected at all. In other words, are those two approaches
calculating the same physical quantity ? Given the fact that one
obtains a retarded amplitude by solving the Dirac equation, while the
perturbative approach calculates a time-ordered amplitude, such an
assumption is highly non-trivial, and in fact very unlikely.

In this paper, we re-examine from first principles the problem of pair
production by two colliding nuclei. Our main focus is on showing that
the two approaches used to study pair production in the peripheral
collision of two ultra-relativistic nuclei are in fact calculating
different physical quantities.  In particular, we give the
relationship between several physical quantities and the appropriate
correlators. The exclusive cross-section of single pair production is
related to a time-ordered (or Feynman) correlator, and is the object
usually calculated in perturbation theory. On the other hand, the
inclusive cross-section of pair production can be related to retarded
amplitudes, which are the ones accessible by solving the Dirac
equation with retarded boundary conditions.  Therefore, the main
discrepancy stems from the fact that the two approaches are
calculating different quantities, despite giving them the same name;
and the two approaches do not have to give the same answer.

We then pursue this analysis by studying the retarded and Feynman
propagators in the background field of two ultra-relativistic nuclei,
and show that these two types of propagators have dramatically
different perturbative expansions: the retarded propagator can be
obtained in closed form in the ultra-relativistic limit, 
as opposed to the Feynman propagator.

We also resolve a puzzle related to the unitarity of the pair
production probabilities calculated in perturbation theory, which
seems to give probabilities larger than 1
\cite{BertuB1,Eichl1,Baur1,Baur2}. We show that this calculation must
include as a corrective factor the vacuum-to-vacuum transition
probability, and that this factor precisely restores unitarity.

The structure of this paper is as follows: after setting up some
notations and details about the model in section \ref{sec:model}, we
recall in section \ref{sec:reduction} some relations between
observable quantities and various types of correlators. We also point
out some uncommon aspects of the Feynman rules that are due to the
fact that the background field is non-static.

We study in section \ref{sec:1-nucleus} the propagators in the
background field of only one nucleus. This case can be solved exactly
in the ultra-relativistic limit, and serves as a starting point for
the more complicated case of two nuclei.
  
In section \ref{sec:watson}, we begin with a formal solution to the
problem of two nuclei, which enables one to write both the retarded
and Feynman propagators in terms of the scattering matrices associated
to the individual nuclei. Then, we explain why the ultra-relativistic
limit simplifies this formula in the case of the retarded propagator,
but does not help for the Feynman one.

In section \ref{sec:unitarity}, we focus on the problem of the
unitarity of the perturbative predictions, and show how it is
resolved. We also show that the multiplicity distribution is not
exactly Poissonian.

Section \ref{sec:approx} discusses two strategies one can use to get
an approximate answer for the Feynman propagator, and hence for the
pair production amplitude.

Finally, section \ref{sec:conclusions} is devoted to concluding
remarks.


\section{Ultra-relativistic limit. Notations}
\label{sec:model}
In this paper, we consider two nuclei of electric charge $Z_1 e$ and
$Z_2 e$ respectively, colliding with an impact parameter $\imb b$. We
study the problem in the frame where the two nuclei have opposite
velocities. For the sake of definiteness, we decide that the nucleus
$Z_1$ is moving in the positive $z$ direction, while the other nucleus
is moving in the negative $z$ direction. We denote by $A_1^\mu(t,{\imb
x})$ and $A_2^\mu(t,{\imb x})$ the electromagnetic vector potentials
created by those nuclei. Because the superposition theorem holds in
QED\footnote{This property is not true in QCD. The problem of
$q\bar{q}$ pair production is therefore complicated by the necessity
of first finding the color vector potential due to the two nuclei.},
the total vector potential is $A^\mu(t,{\imb x})=A_1^\mu(t,{\imb
x})+A_2^\mu(t,{\imb x})$.

In the ultra-relativistic limit, we can assume that the two nuclei do
not recoil, and simply treat them as a classical background field.
The trajectories of the two nuclei are respectively $z=\pm t$ ($c=1$),
${\imb x_\perp}=\pm{\imb b}/2$, and there is a gauge in which
$A_1^\mu$ and $A_2^\mu$ have the following form
\cite{BaltzRW1,Baltz1,Baltz2}:
\begin{eqnarray}
&&A_1^\mu(t,{\imb x})=Z_1 e\delta(v_+\cdot x) v_+^\mu 
\ln\left({{({\imb x_\perp}-{\imb b}/2)^2}\over{{\imb b}^2}}\right)
\; ,\nonumber\\
&&A_2^\mu(t,{\imb x})=Z_2 e\delta(v_-\cdot x) v_-^\mu 
\ln\left({{({\imb x_\perp}+{\imb b}/2)^2}\over{{\imb b}^2}}\right)
\; ,
\label{eq:ur-potentials}
\end{eqnarray}
where $v_\pm^\mu\equiv(1,0,0,\pm 1)/\sqrt{2}$. This choice of gauge
makes the following fact obvious: the electromagnetic field of an
ultra-relativistic nucleus is confined in its transverse plane due to
Lorentz contraction. In the ultra-relativistic approximation, we
neglect any modification of the trajectories of the two nuclei.
Therefore, one can see this problem as a field theory for fermions in
a classical electromagnetic background field.  An essential property
of the field generated by two colliding nuclei is that there is no
frame in which it is time-independent (static).  This property has
important consequences on the field theory describing fermions in this
background; one of them is that pairs can be created\footnote{This is
  to be contrasted with the case of a single nucleus. In the frame of
  that nucleus, the background field is static, and it is well know
  that a single nucleus moving on a straight line does not produce pairs
  (except via the non-perturbative Schwinger mechanism of vacuum
  instability \cite{Schwi2}, which we ignore here).}.

However, one should realize that the limit of infinite momentum for
the nuclei is only an approximation of the real physical situation.
Indeed, the potentials of Eq.~(\ref{eq:ur-potentials}) cannot be
expected to describe the motion of a lepton comoving with one of the
nuclei.  This restriction also appears in the work of
\cite{JackiKO1,WellsSE1}.  In particular, \cite{WellsSE1} shows that
the gauge transformation that leads to Eq.~(\ref{eq:ur-potentials}) is
well-defined only if the lepton is not comoving with one of the
nuclei.  There is another way to see the problem with
Eq.~(\ref{eq:ur-potentials}): this background potential is
short-ranged in the $z$ and $t$ directions, while the potential before
the gauge transformation was long-ranged. The absence of interaction
at asymptotic times allows the construction of plane-wave asymptotic
states, and makes the usual formulation of reduction theory
applicable. But it is also clear that those interactions cannot be
removed by a gauge transformation for a comoving lepton, the states of
which are described by distorted waves. In conclusion, those
potentials are valid only if the typical interaction time between the
lepton and the nuclei goes like $1/\gamma$, i.e. for leptons produced
in the mid-rapidity region.

In addition, the description of the background field by a delta
function is probably not accurate at large transverse distances if the
Lorentz factor $\gamma$ is finite. Indeed, at large distances, the
lines of force of the electric field should have some curvature and
depart from the transverse plane.  Therefore,
Eq.~(\ref{eq:ur-potentials}) is expected to break down also at large
impact parameter ${\imb b}$ or if the transverse separation between
the lepton and a nucleus is large. This problem is expected to show up
via infrared divergences in the transverse momentum integrals. In
\cite{BaltzM1}, it was argued that those divergences are regularized
by an effective cutoff of order $\omega/\gamma$. Although this
argument cannot tell precisely what this cutoff should be, it is
obvious from the derivation of the ultra-relativistic limit of the
potentials that it is completely determined by the kinematics and is
therefore independent of $Z_{1,2}\alpha$ (the atomic numbers and
coupling constants appear only in the numerator of the Coulomb
potential). Therefore, the uncertainty in the cutoff can affect only
the overall normalization of integrated cross-sections, but not its
dependence on atomic numbers.

In view of those caveats, one should take with a grain of salt the
results obtained in this model for completely integrated (over the
pair phase-space and over impact parameter) cross-sections, because
there are regions in such an integral where the above approximation may
not be appropriate.

One should realize that due to the large electric charge $Ze$ of a
nucleus like gold, the addition of an extra photon connecting the
electron line to a nucleus brings a factor $Z\alpha$ ($\alpha\equiv
e^2/4\pi$), which may be of order $1$. Therefore, we want to include
as much as possible of these ``perturbative'' corrections. On the
contrary, photons coupling an electron line to itself (or to another
electron line) yield only a factor of $\alpha$, which is indeed a
small correction. Therefore, we also neglect the interactions of the
electrons with dynamical photons, and keep only the classical photon
background. The Lagrangian density for the electron field $\psi(x)$ is
therefore
\begin{equation}
{\cal L}\equiv \overline{\psi}(x)(i\slpartial_x-e\,\slA(x)-m)\psi(x)
\label{eq:Lagrangian}
\end{equation}
in this model.

In the following, we make extensive use of the light-cone coordinates.
For any 4-vector $x^\mu$, we define:
\begin{equation}
x^\pm\equiv {{x^0\pm x^3}\over{\sqrt{2}}}\; ,
\end{equation}
and denote by ${\imb x_\perp}$ the transverse part of the 3-vector
$\imb x$. With these notations, the invariant norm of $x^\mu$ is
$x^2=2x^+ x^- -{\imb x_\perp}^2$, and the scalar product of $k^\mu$
and $x^\mu$ is $k\cdot x=k^+ x^-+k^- x^+ -{\imb k_\perp}\cdot{\imb
  x_\perp}$. The invariant measure $d^4x$ becomes
$d^4x=dx^+dx^-d^2{\imb x_\perp}$.  Note also that $x^\pm= x\cdot
v_\mp$ with the $v_\pm$ defined above.

\section{Reduction formulae and Feynman rules}
\label{sec:reduction}
In this section, we relate observable quantities to correlation
functions of the fermionic field operator. In fact, these
considerations do not depend on the nature of the background field,
and rely only on the Lagrangian of Eq.~(\ref{eq:Lagrangian}). In
particular, the formulae of this section are independent of the
ultra-relativistic approximation.
\subsection{Amplitude to produce one pair}
\subsubsection{Reduction formula}
The amplitude to produce one $e^+e^-$ pair is
\begin{equation}
M_1({\imb p},{\imb q})\equiv 
\left<e^+({\imb p})e^-({\imb q}){}_{\rm out}|0_{\rm in}\right>= 
\left<0_{\rm out}|d_{\rm out}({\imb p})b_{\rm out}({\imb q})|0_{\rm in}\right>\; ,
\end{equation}
where $d^\dagger_{\rm out}({\imb p})$ (resp. $b^\dagger_{\rm
  out}({\imb p})$ ) is the operator that creates a positron (resp. an
electron) of 3-momentum $\imb p$ in the final state. At this stage, it
is very important to carefully distinguish the in- and out- states and
operators\footnote{For instance, $b_{\rm out}$ does not annihilate the
  in-vacuum: $\left.b_{\rm out}|0_{\rm in}\right>\not=0$, while
  $\left.b_{\rm out}|0_{\rm out}\right>=0$. It is precisely this fact
  that makes the pair production amplitude non-zero in a non-static
  background field.}.
 
Making use of the following relations between the
annihilation/creation operators\footnote{For the sake of brevity, we
do not write explicitly the spin indices on spinors and
creation/annihilation operators. In formulae for cross-sections, it is
implicitly assumed that we sum over the spin of the final state
particles.} and the field itself (see
\cite{ItzykZ1}, page 61):
\begin{eqnarray}
&&b_{\rm out}({\imb q})=\int d^3{\imb x}\; \overline{u}({\imb q}) \gamma^0 \psi_{\rm out}(t,{\imb x}) e^{iq\cdot x}\; ,\nonumber\\
&&d_{\rm out}({\imb p})=\int d^3{\imb x}\; \overline{\psi}_{\rm out}(t,{\imb x})\gamma^0 v({\imb p}) e^{ip\cdot x}\; ,
\end{eqnarray}
where $p_0$ and $q_0$ are, respectively, $\surd({{\imb p}^2}+m^2)$ and
$\surd({{\imb q}^2}+m^2)$ (the time $x^0=t$ that shows up in these
formulae is irrelevant, and disappears in the course of the
calculation), one can show by standard manipulations (\cite{ItzykZ1},
pages 205-207) that the pair production amplitude defined above can be
related to a 2-point time-ordered correlator by the following reduction
formula:
\begin{eqnarray}
&&\left<0_{\rm out}|d_{\rm out}({\imb p})
b_{\rm out}({\imb q})|0_{\rm in}\right>
=\left[{i\over{\sqrt{{\cal Z}_2}}}\right]^2\int d^4x\, d^4y\;\nonumber\\
&&\qquad\times e^{iq\cdot x}
\overline{u}({\imb q})(i\stackrel{\rightarrow}{\slpartial}_x-m)
\left<0_{\rm out}|{\rm T}\overline{\psi}(y)\psi(x)|0_{\rm in}
\right>(i\stackrel{\leftarrow}{\slpartial}_y+m)v({\imb p}) e^{ip\cdot y}\; ,
\label{eq:reduction}
\end{eqnarray}
where the arrows indicate on which side the derivatives act
(preventing their action on the exponentials), and where ${\cal Z}_2$
is the wave function renormalization factor for an
electron\footnote{With the Lagrangian of Eq.~(\ref{eq:Lagrangian}),
  this factor is equal to $1$.}. The important point to note here is
that the average value of the time-ordered product is taken between
the in- and out- vacua, which are {\sl different} states. To expand a
little on this, let us add that we necessarily have $|\left<0_{\rm
    out}|0_{\rm in}\right>|^2<1$ in background fields that can produce
pairs, because of unitarity (see section \ref{sec:unitarity} for the
role played by $\left<0_{\rm out}|0_{\rm in}\right>$ in issues related
to unitarity).

\subsubsection{Perturbative expansion of $\left<0_{\rm out}|{\rm T}\,\overline{\psi}(y)\psi(x)|0_{\rm in}\right>$ }
We need to calculate the correlator $\left<0_{\rm out}|{\rm
    T}\,\overline{\psi}(y)\psi(x)|0_{\rm in}\right>$. From that,
Eq.~(\ref{eq:reduction}) tells us how to obtain the pair production
amplitude: amputate the external legs of the correlator, take its
Fourier transform, and insert the result between the appropriate
spinors.

The Feynman rules to calculate perturbatively this correlator can be
obtained by switching to the interaction picture. The Heisenberg field
can be expressed in terms of the field in the interaction
representation, via the following relation:
\begin{equation}
\psi(x)\equiv U(t_{_{I}},x^0)\psi_{_{I}}(x)U(x^0,t_{_{I}})\; ,
\end{equation}
where $t_{_{I}}$ is the time at which the Heisenberg and interaction
pictures coincide (ultimately, we will take $t_{_{I}}$ to $-\infty$),
and where the evolution operator $U$ is related to the interaction
part ${\cal L}_{\rm int}\equiv -e\overline{\psi}(x)\slA(x)\psi(x)$ of
the Lagrangian by
\begin{equation}
U(t_2,t_1)={\rm P}_{12}\exp i\int_{t_1}^{t_2}d^4x\;{\cal L}_{\rm int}(\psi_{_{I}}(x))\; ,
\end{equation}
where ${\rm P}_{12}$ is an ordering operator along the path connecting
$t_1$ to $t_2$ (ordinary time-ordering if $t_1<t_2$, and reverse
time-ordering if $t_1>t_2$).  Using this transformation, and taking
$t_{_{I}}\to -\infty$ (in this limit, $\psi_{_{I}}\to \psi_{\rm in}$),
we have
\begin{eqnarray}
&&\left<0_{\rm out}|{\rm T}\overline{\psi}(y)\psi(x)|0_{\rm in}\right>
=\Big<0_{\rm out}\Big|U(-\infty,+\infty){\rm T}\overline{\psi}_{\rm in}(y)\psi_{\rm in}(x)\nonumber\\
&&\qquad\qquad\qquad\qquad\qquad\qquad\times\exp i\int_{-\infty}^{+\infty}d^4x {\cal L}_{\rm int}(\psi_{\rm in}(x))
\Big|0_{\rm in}\Big>\; .
\label{eq:pert-exp}
\end{eqnarray}
Noticing that $\left<0_{\rm
    out}|U(-\infty,+\infty)\right.=\left<0_{\rm in}|\right.$, we have
simply:
\begin{equation}
\left<0_{\rm out}|{\rm T}\overline{\psi}(y)\psi(x)|0_{\rm in}\right>
=\Big<0_{\rm in}\Big|{\rm T}\overline{\psi}_{\rm in}(y)\psi_{\rm in}(x)
\exp i\int_{-\infty}^{+\infty}d^4x {\cal L}_{\rm int}(\psi_{\rm in}(x))
\Big|0_{\rm in}\Big>\; .
\end{equation}
Everything being now expressed in terms of in-fields and in-states,
the right hand side of the last equation can be evaluated by the
standard perturbative expansion of the exponential.

This perturbative expansion has, however, a peculiarity due to the
fact that the background field is non-static: the vacuum-vacuum
diagrams (i.e. diagrams without any external legs) do not cancel, and
their sum is not a phase. On the contrary, in a conventional field
theory, $\left<0_{\rm out}|0_{\rm in}\right>$ is just an irrelevant
phase, and one takes advantage of this fact to divide the r.h.s. of
the previous equation by $\left<0_{\rm in}|U(+\infty,-\infty)|0_{\rm
    in}\right>$. Then, one can show that this denominator cancels
(\cite{ItzykZ1}, pages 266-267) order by order the vacuum-vacuum
diagrams\footnote{This step is not mandatory though.  One can live
  with the vacuum-vacuum diagrams and notice that they add up to a
  pure phase, so that they always drop out in the calculation of cross
  sections, even if they are present in the amplitude.}. This trick
cannot be used here due to the fact that $\left<0_{\rm out}|0_{\rm
    in}\right>$ is not a phase in the present problem \cite{Feynm1}.
In fact, this is deeply rooted in the property that the background can
produce particles.

We can keep this complication aside for a while by just writing:
\begin{eqnarray}
&&\left<0_{\rm out}|{\rm T}\overline{\psi}(y)\psi(x)|0_{\rm in}\right>
=\left<0_{\rm in}|U(+\infty,-\infty)|0_{\rm in}\right>\nonumber\\
&&\qquad\qquad\times{{\left<0_{\rm in}\Big|{\rm T}\overline{\psi}_{\rm in}(y)\psi_{\rm in}(x)
\exp i\int_{-\infty}^{+\infty}d^4x {\cal L}_{\rm int}(\psi_{\rm in}(x))
\Big|0_{\rm in}\right>}\over{\left<0_{\rm in}|U(+\infty,-\infty)|0_{\rm in}\right>}}\; ,
\label{eq:GF-trick}
\end{eqnarray}
so that the fraction on the right hand side has a perturbative
expansion where the vacuum-vacuum diagrams do cancel. In fact, this
fraction is nothing but the Feynman propagator
\begin{equation}
G_F(x,y)\equiv {{\left<0_{\rm out}|{\rm T}\overline{\psi}(y)\psi(x)
|0_{\rm in}\right>}\over{\left<0_{\rm out}|0_{\rm in}\right>}}
\end{equation}
 of an electron in the
electromagnetic background field. Its perturbative expansion is the
usual one: it is obtained by inserting the external potential
$-ie\slA(x)$ on chains of free Feynman propagators
$G_{_{F}}^0(x,y)\equiv\left<0_{\rm in}|{\rm T}\overline{\psi}_{\rm
    in}(y)\psi_{\rm in}(x)|0_{\rm in}\right>$.  In particular, the
perturbative expansion for the full propagator can be generated by the
following Lippmann-Schwinger equation:
\begin{equation}
G(x,y)=G^0(x,y)-ie\int d^4z\; G^0(x,z)\,\slA(z) G(z,y)\; .
\end{equation}
Note that this equation is equally valid for the Feynman and for the
retarded propagator (provided the free propagator $G^0$ is chosen
accordingly). This remark will become important later when we also
need to study the retarded propagator.

It turns out to be more convenient to work in Fourier space. If we
define\footnote{We do not use a distinct symbol for Fourier
  transforms, as the context always enables to tell the difference.}
\begin{equation}
G(x,y)\equiv\int {{d^4p}\over{(2\pi)^4}} {{d^4q}\over{(2\pi)^4}} e^{-iq\cdot x}
e^{ip\cdot y} G(q,p)\; ,
\end{equation}
then the Lippmann-Schwinger equation becomes
\begin{equation}
G(q,p)=(2\pi)^4\delta(p-q)G^0(p)-ieG^0(q)\int{{d^4k}\over{(2\pi)^4}}\,\slA(k)
G(q+k,p)\; ,
\end{equation} 
with $G^0(p)\equiv i/(\,\slp-m)$ (and an unspecified $i\epsilon$
prescription depending on whether we are studying the retarded or the
Feynman propagator) and
\begin{equation}
A^\mu(-k)\equiv\int d^4x\;e^{ik\cdot x} A^\mu(x)\; .
\end{equation}

If one introduces the interacting part ${\cal T}_F$ of the Feynman
propagator by the relation
\begin{equation}
G_F(q,p)=
(2\pi)^4\delta(p-q)G^0_F(p)+G^0_F(q){\cal T}_F(q,p)G^0_F(p)\; ,
\end{equation}
then Eqs.~(\ref{eq:reduction}) and (\ref{eq:GF-trick}) lead to the
following expression for the probability to produce exactly one pair
in a collision at impact parameter ${\imb b}$:
\begin{equation}
P_1=|\left<0_{\rm out}|0_{\rm in}\right>|^2
\int{{d^3 {\imb q}}\over{(2\pi)^3 2\omega_{\imb q}}}
\int{{d^3 {\imb p}}\over{(2\pi)^3 2\omega_{\imb p}}}
\Big|
\overline{u}({\imb q}){\cal T}_F(q,-p)v({\imb p})
\Big|^2\; .
\label{eq:P1}
\end{equation}
Note that this formula has been derived independently of the
details of the background field, and is therefore completely general.

\subsection{Average number of produced pairs}
\subsubsection{Expression as a correlator}
The approach based on solving the Dirac equation was motivated by the
papers \cite{ReinhMGS1,RumriMSGG1,WellsOUBS1}, where a formula giving
the average number of pairs produced per collision is derived in terms
of retarded amplitudes only. We present here a justification of this
formula in the field-theoretical framework we are following in this
paper.

Let us start from the expression of the average number of pairs
$\overline{n}$ as the sum
\begin{equation}
\overline{n}=\sum_{n=1}^{+\infty}n P_n=\sum_{n=0}^{+\infty} (n+1)P_{n+1}\; ,
\end{equation}
where we denote by $P_n$ the probability to produce exactly $n$ pairs
in a collision at impact parameter ${\imb b}$. Making explicit what
this probability is, we find
\begin{eqnarray}
\overline{n}&&=\sum_{n=0}^{+\infty}(n+1) {1\over{(n+1)!^2}}
\int\prod_{i=1}^{n+1} {{d^3{\imb p}_i}\over{(2\pi)^3 2\omega_{{\imb p}_i}}}
{{d^3{\imb q}_i}\over{(2\pi)^3 2\omega_{{\imb q}_i}}}\nonumber\\
&&\qquad\times
\Big|\Big<0_{\rm out}\Big|d_{\rm out}({\imb p}_1)b_{\rm out}({\imb q}_1)\cdots
d_{\rm out}({\imb p}_{n+1})b_{\rm out}({\imb q}_{n+1})
\Big|0_{\rm in}\Big>\Big|^2\; ,
\end{eqnarray}
where we denote $\omega_{\imb p}\equiv \surd{({{\imb p}^2}+m^2)}$.
Singling out one of the electrons (say the one with momentum ${\imb
  q}_{n+1}$, which we call simply ${\imb q}$) and expanding the
squared modulus, we can write:
\begin{eqnarray}
\overline{n}&&=\int{{d^3{\imb q}}\over{(2\pi)^3 2\omega_{\imb q}}}
\Big<0_{\rm in}\Big|b^\dagger_{\rm out}({\imb q})\nonumber\\
&&\qquad\times\sum_{n=0}^{+\infty}{1\over{n!}}{1\over{(n+1)!}}
\int\prod_{i=1}^{n+1} {{d^3{\imb p}_i}\over{(2\pi)^3 2\omega_{{\imb p}_i}}}
\prod_{j=1}^{n} {{d^3{\imb q}_j}\over{(2\pi)^3 2\omega_{{\imb q}_j}}}
\nonumber\\
&&\qquad\quad
\Big|d^\dagger_{\rm out}({\imb p}_1)b^\dagger_{\rm out}({\imb q}_1)\cdots
d^\dagger_{\rm out}({\imb p}_{n})b^\dagger_{\rm out}({\imb q}_{n})
d^\dagger_{\rm out}({\imb p}_{n+1})\Big|0_{\rm out}\Big>
\nonumber\\
&&\qquad\qquad
\Big<0_{\rm out}\Big|
d_{\rm out}({\imb p}_1)b_{\rm out}({\imb q}_1)\cdots
d_{\rm out}({\imb p}_{n})b_{\rm out}({\imb q}_{n})
d_{\rm out}({\imb p}_{n+1})\Big|\nonumber\\
&&\qquad\times b_{\rm out}({\imb q})\Big|0_{\rm in}\Big>\; .
\label{eq:nbar-inter}
\end{eqnarray}
Noticing now that the three intermediate lines are the identity
operator on the subspace of states with electric charge $+e$, we find:
\begin{equation}
\overline{n}=\int{{d^3{\imb q}}\over{(2\pi)^3 2\omega_{\imb q}}}
\Big<0_{\rm in}\Big|b^\dagger_{\rm out}({\imb q})b_{\rm out}({\imb q})\Big|
0_{\rm in}\Big>\; .
\end{equation}
This formula simply tells that in order to count the number of pairs
produced if the initial state is the vacuum, it is sufficient to count
the number of electrons in the final state. Had we decided to single
out a positron annihilation operator in Eq.~(\ref{eq:nbar-inter}), we
would have obtained instead:
\begin{equation}
\overline{n}=\int{{d^3{\imb p}}\over{(2\pi)^3 2\omega_{\imb p}}}
\Big<0_{\rm in}\Big|d^\dagger_{\rm out}({\imb p})d_{\rm out}({\imb p})\Big|
0_{\rm in}\Big>\; .
\end{equation}
One can note that if the background potential has time-reversal
symmetry, we have:
\begin{equation}
\overline{n}=\!\!\int\!\!{{d^3{\imb q}}\over{(2\pi)^3 2\omega_{\imb q}}}
\Big<0_{\rm out}\Big|b^\dagger_{\rm in}({\imb q})b_{\rm in}({\imb q})\Big|
0_{\rm out}\Big>=\!\!\int\!\!{{d^3{\imb p}}\over{(2\pi)^3 2\omega_{\imb p}}}
\Big<0_{\rm out}\Big|d^\dagger_{\rm in}({\imb p})d_{\rm in}({\imb p})\Big|
0_{\rm out}\Big>\; .
\end{equation}

There are also reduction formulae for those correlators, that give for
instance
\begin{eqnarray}
\overline{n}&&=\int{{d^3{\imb q}}\over{(2\pi)^3 2\omega_{\imb q}}}
\left[{i\over{\sqrt{{\cal Z}_2}}}\right]^2\int d^4x\, d^4y\;\nonumber\\
&&\quad\times e^{iq\cdot x}
\overline{u}({\imb q})(i\!\stackrel{\rightarrow}{\slpartial}_x-m)
\left<0_{\rm in}|\overline{\psi}(y)\psi(x)|0_{\rm in}
\right>(i\!\stackrel{\leftarrow}{\slpartial}_y-m)u({\imb q}) e^{-iq\cdot y}\; .
\label{n-bar-reduction}
\end{eqnarray}
Therefore, in order to calculate the average number of pairs, we need
now the ordinary product of two fields, averaged with the initial
vacuum $\left|0_{\rm in}\right>$ on both sides.

\subsubsection{Perturbative expansion of $\left<0_{\rm in}|\overline{\psi}(y)\psi(x)|0_{\rm in}\right>$}
\label{sec:keldysh}
In order to switch to the interaction representation, it is easier to
start with $\left<0_{\rm in}|{\rm T}\overline{\psi}(y)\psi(x)|0_{\rm
    in}\right>$ for which one can write directly
\begin{eqnarray}
&&\left<0_{\rm in}|{\rm T}\overline{\psi}(y)\psi(x)|0_{\rm in}\right>
=\Big<0_{\rm in}\Big|U(-\infty,+\infty){\rm T}\overline{\psi}_{\rm in}(y)\psi_{\rm in}(x)\nonumber\\
&&\qquad\qquad\qquad\qquad\qquad\qquad\times\exp i\int_{-\infty}^{+\infty}d^4x {\cal L}_{\rm int}(\psi_{\rm in}(x))
\Big|0_{\rm in}\Big>\; .
\label{eq:pert-exp-in-in}
\end{eqnarray}
This time, one cannot get rid of the $U(-\infty,+\infty)$ in the right
hand side, but there is a standard trick to incorporate it in the
perturbative expansion \cite{Keldy1,Schwi1,BakshM1}. For that, one has
to introduce a contour ${\cal C}$ going from $-\infty$ to $+\infty$
just above the real-time axis, and then back from $+\infty$ to
$-\infty$ below the real axis:
\setbox1=\hbox to 9cm{\resizebox*{9cm}{!}{\includegraphics{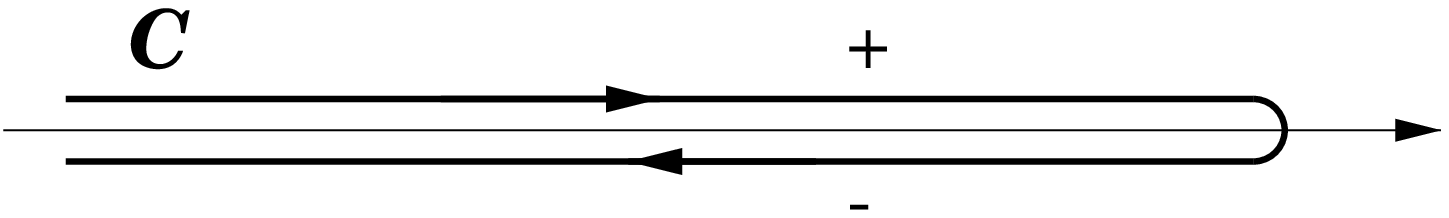}}}
\begin{displaymath}
\raise -11mm\box1
\end{displaymath}
One can then check that
\begin{equation}
\left<0_{\rm in}|{\rm T}\overline{\psi}(y)\psi(x)|0_{\rm in}\right>
=\Big<0_{\rm in}\Big|{\rm P}\overline{\psi}_{\rm in}(y)\psi_{\rm in}(x)\exp i\int_{\cal C}d^4x {\cal L}_{\rm int}(\psi_{\rm in}(x))
\Big|0_{\rm in}\Big>\; .
\end{equation}
In the previous formula, $\rm P$ stands for an ordering of operators
along the path ${\cal C}$, identical to the usual time-ordering $\rm
T$ on the upper branch of the contour (and with the convention that
points on the lower branch have a ``larger'' time than points on the
upper branch).  We therefore have formally similar Feynman rules for
the perturbative expansion of $\left<0_{\rm in}|{\rm
    T}\overline{\psi}(y)\psi(x)|0_{\rm in}\right>$, except that the
time integrations at each insertion of the external potential are
performed on the contour ${\cal C}$.

It is customary to write this formalism in matrix form
\cite{Keldy1,LandsW1,Bella1} (in this section, objects denoted by a boldface
letter are $2\times 2$ matrices), by splitting the free propagator in
four components according to where $x^0$ and $y^0$ lie on ${\cal C}$
(upper or lower branch):
\begin{equation}
{\imb G}^0(x,y)\equiv \pmatrix{
G_{++}^0(x,y) &G_{+-}^0(x,y) \cr
G_{-+}^0(x,y) &G_{--}^0(x,y) \cr
}\; ,
\end{equation}
with
\begin{eqnarray}
&&G_{++}^0(x,y)\equiv\left<0_{\rm in}|
{\rm T}\overline{\psi}_{\rm in}(y)\psi_{\rm in}(x)|0_{\rm in}\right>\; ,
\nonumber\\
&&
G_{--}^0(x,y)\equiv \left<0_{\rm in}|\overline{\rm T}\,
\overline{\psi}_{\rm in}(y)\psi_{\rm in}(x)|0_{\rm in}\right>\; ,
\nonumber\\
&&G_{+-}^0(x,y)\equiv\left<0_{\rm in}|
\overline{\psi}_{\rm in}(y)\psi_{\rm in}(x)|0_{\rm in}\right>\; ,
\nonumber\\
&&
G_{-+}^0(x,y)\equiv\left<0_{\rm in}|
-\psi_{\rm in}(x)\overline{\psi}_{\rm in}(y)|0_{\rm in}\right>\; ,
\end{eqnarray}
where $\overline{\rm T}$ is the reverse time-ordering operator. The
correlator $\left<0_{\rm in}|\overline{\psi}(y)\psi(x)|0_{\rm
    in}\right>$ we need for the average number of produced electrons
is the $+-$ component of the exact matrix propagator.

One needs also to give a matrix structure to the external potential
$\slA(x)$:
\begin{equation}
\slbfA(x)\equiv \tau_3\, \slA(x)\; ,
\end{equation}
where $\tau_3\equiv{\rm Diag}(1,-1)$ is the third Pauli
matrix\footnote{The $-$ sign for the $--$ component of $\slbfA$ comes
  from the fact that the integral over the lower branch of the time
  contour goes from $+\infty$ back to $-\infty$.}.  Then, at each
vertex, the rule is to integrate only over the ordinary time axis, and
to multiply the matrices corresponding to the propagators and
potential insertions (in the order they appear in the Feynman
diagram). For instance, the Lippmann-Schwinger equation for the exact
matrix propagator in Fourier space is
\begin{equation}
{\imb G}(q,p)=(2\pi)^4\delta(p-q){\imb G}^0(p)-ie{\imb G}^0(q)\int{{d^4k}\over{(2\pi)^4}}\,\slbfA(k)
{\imb G}(q+k,p)\; .
\label{eq:lippmann-matrix}
\end{equation}
Note that in Fourier space, the expression of the free matrix
propagator is\footnote{The expression of those propagators can be
  found in \cite{Bella1} (Eqs.~(3.93)). To apply them here, the
  following substitutions must be made: $\sigma=0$, $n(p_0)=0$. Also,
  \cite{Bella1} is using different notations for the
  indices labelling the two branches of the contour: $1\equiv+$,
  $2\equiv -$. }:
\begin{eqnarray}
&&G_{++}^0(p)=i{{\slp+m}\over{p^2-m^2+i\epsilon}}=[G_{--}^0(p)]^*\; ,
\nonumber\\
&&G_{+-}^0(p)=2\pi\theta(- p_0)(\slp+m)\delta(p^2-m^2)\; ,
\nonumber\\
&&G_{-+}^0(p)=2\pi\theta(+ p_0)(\slp+m)\delta(p^2-m^2)\; .
\end{eqnarray}

As it stands, the expansion of Eq.~(\ref{eq:lippmann-matrix}) has a
very intricate matrix structure. However, it can be simplified by
applying a ``rotation'' \cite{AurenB1,EijckW1,EijckKW1} on the
previously defined matrices. Let us define
\begin{eqnarray}
&&{\imb G}_U^0(p)\equiv U(p) {\imb G}^0(p) U^{^{T}}(-p)\; ,\nonumber\\
&&{\imb G}_U(q,p)\equiv U(q){\imb G}(q,p) U^{^{T}}(-p)\; ,\nonumber\\
&&\slbfA_U(q,q+k)\equiv U^{^{T}}{}^{-1}(-q)\,\slbfA(k) U^{-1}(q+k)\; ,
\end{eqnarray}
where $U$ is an invertible matrix. The Lippmann-Schwinger
equation hardly changes in the rotated formalism,
\begin{equation}
{\imb G}_U(q,p)=(2\pi)^4\delta(p-q){\imb G}_U^0(p)-ie{\imb G}_U^0(q)\int{{d^4k}\over{(2\pi)^4}}\,\slbfA_U(q,q+k)
{\imb G}_U(q+k,p)\; ,
\label{eq:lippmann-rotated}
\end{equation}
but there are some choices of $U$ that simplify significantly the free
matrix propagator.  A convenient choice is
\begin{equation}
U(p)\equiv {1\over{\sqrt{2}}}\pmatrix{
1 & -1\cr
1 & 1\cr
}\; ,
\label{eq:RA}
\end{equation}
which leads to the free propagator
\begin{equation}
{\imb G}_U^0(p)=\pmatrix{
0 & G_A^0(p)\cr
G_R^0(p) & G_S^0(p)\cr
}\; ,
\end{equation}
with the definitions (the index ``$S$'' stands for ``on-shell''):
\begin{eqnarray}
&&G_R^0(p)\equiv i {{\slp+m}\over{p^2-m^2+ip_0\epsilon}}\;,\qquad
G_A^0(p)\equiv i {{\slp+m}\over{p^2-m^2-ip_0\epsilon}}\nonumber\\
&&G_S^0(p)\equiv 2\pi(\slp+m)\delta(p^2-m^2)\; .
\end{eqnarray}
In this transformation, the external potential becomes
\begin{equation}
\slbfA_U(q,q+k)=\slA(k)\,\pmatrix{
0 & 1\cr
1 & 0 \cr
}\; .
\end{equation}
Therefore, we have a simplification in the propagator which has now a
vanishing component, the non-zero components being the free retarded
and advanced propagators and the very simple on-shell piece $G_S^0$.
That this transformation helps to resum the perturbative expansion is
readily seen by computing the building block that will be iterated in
the expansion:
\begin{equation}
\slbfA_U(q,q+k){\imb G}_U^0(q+k)\!=\!\slA(k)\left[
\pmatrix{
G_R^0(q+k) & 0\cr
0 & \!\!\!\!\!\!\!\!\!G_A^0(q+k)\cr
}
\!+\!
G_S^0(q+k)\pmatrix{0 & 1\cr
0 & 0\cr}
\right]\; .
\end{equation}
Indeed, the fact that this object is the sum of a diagonal matrix and
a nilpotent matrix leads trivially to the following formula for a term
with $n$ insertions of $\slbfA$:
\begin{eqnarray}
&&{\imb G}_U^0(q_0)\Big[\,\slbfA_U(q_0,q_1){\imb G}_U^0(q_1)\Big]
\cdots \Big[\,\slbfA_U(q_{n-1},q_n){\imb G}_U^0(q_n)\Big]=\nonumber\\
&&=
\pmatrix{
\qquad 0 \qquad\quad
G_A^0(q_0)\,\slA(q_1-q_0)G_A^0(q_1)\cdots \slA(q_n-q_{n-1})G_A^0(q_n)
\vphantom{\Big[}\cr
G_R^0(q_0)\,\slA(q_1-q_0)G_R^0(q_1)\cdots \slA(q_n-q_{n-1})G_R^0(q_n)
\qquad\quad 0\qquad\cr
}\nonumber\\
&&+
\sum_{i=0}^n G_R^0(q_0)\,\slA(q_1-q_0)\cdots G_R^0(q_{i-1})
\,\slA(q_i-q_{i-1})G_S^0(q_i)\nonumber\\[-5mm]
&&\qquad\quad\times
{\slA(q_{i+1}-q_i) 
G_A^0(q_{i+1})\,\slA(q_{i+2}-q_{i+1})\cdots G_A^0(q_n)
\pmatrix{
0 & 0\cr
0 & 1\cr
}\; .}
\end{eqnarray}
At this stage, it is straightforward to sum over the number $n$ of
insertions of the external potential, to obtain the exact propagator
in the new basis (set $q\equiv q_0$ and $p\equiv q_n$ in the previous
formula):
\begin{equation}
{\imb G}_U(q,p)=\pmatrix{
0 & G_A(q,p) \cr
G_R(q,p) & G_S(q,p)\cr
}\; ,
\label{eq:RA-result}
\end{equation}
where $G_{R,A}(q,p)$ are the exact retarded and advanced propagators,
and where
\begin{eqnarray}
G_S(q,p)&&\equiv \int {{d^4k}\over{(2\pi)^4}}2\pi\delta(k^2-m^2)
\nonumber\\
&&\qquad\times
G_R(q,k)G_R^0{}^{-1}(k)(\slk+m)G_A^0{}^{-1}(k)
G_A(k,p)\; .
\label{eq:Gt-1}
\end{eqnarray}
At this stage, we have an exact expression for the Fourier transform
$G_{+-}(q,p)$ of $\left<0_{\rm in}|\overline{\psi}(y)\psi(x)|0_{\rm
    in}\right>$, which reads
\begin{equation}
G_{+-}(q,p)
={1\over 2}\Big[G_A(q,p)-G_R(q,p)+G_S(q,p)\Big]\; .
\label{eq:approx-1}
\end{equation}
Using Eq.~(\ref{eq:Gt-1}), we see that this correlator can be written
in closed form in terms of $G_R$ and $G_A$\footnote{It is now clear
why Eq.~(\ref{eq:lippmann-matrix}) had a very complicated matrix
structure: trying to solve directly this matrix equation amounts to
write the exact retarded and advanced propagators in terms of bare
time-ordered propagators, which is not easy in practice. This explains
why the rotation leading to the retarded/advanced basis simplified a
lot this calculation.}. According to Eq.~(\ref{n-bar-reduction}),
$\overline{n}$ is obtained by amputating the external legs of
$G_{+-}(q,q)$. To that effect, it is convenient to introduce the
interacting parts ${\cal T}_R(q,p)$ and ${\cal T}_A(q,p)$ of the exact
retarded and advanced propagators, defined by
\begin{eqnarray}
&&G_R(q,p)=
(2\pi)^4\delta(p-q)G^0_R(p)+G^0_R(q){\cal T}_R(q,p)G^0_R(p)\nonumber\; ,\\
&&G_A(q,p)=
(2\pi)^4\delta(p-q)G^0_A(p)+G^0_A(q){\cal T}_A(q,p)G^0_A(p)\; ,
\end{eqnarray}
which leads to 
\begin{eqnarray}
&&\overline{n}={1\over 2}\int{{d^3{\imb q}}\over{(2\pi)^2 2\omega_q}}
\overline{u}({\imb q})\Big[{\cal T}_A(q,q)-{\cal T}_R(q,q)\nonumber\\
&&\qquad+
\int {{d^4k}\over{(2\pi)^4}}2\pi\delta(k^2-m^2)
{\cal T}_R(q,k)(\slk+m){\cal T}_A(k,q)
\Big]u({\imb q})\; .
\end{eqnarray}
By using the Lippmann-Schwinger equations for the retarded and advanced
propagators, one can check that\footnote{This relation can be seen as
  a form of the optical theorem.}
\begin{eqnarray}
{\cal T}_R(q,p)-{\cal T}_A(q,p)&&=
\int{{d^4k}\over{(2\pi)^4}}2\pi[\theta(k_0)-\theta(-k_0)]\delta(k^2-m^2)
\nonumber\\
&&\qquad\times{\cal T}_R(q,k)(\slk+m){\cal T}_A(k,p)\; .
\end{eqnarray}
From there, we can further simplify $\overline{n}$, and obtain:
\begin{eqnarray}
\overline{n}&&=\int{{d^3{\imb q}}\over{(2\pi)^2 2\omega_q}}
\int {{d^4k}\over{(2\pi)^4}}2\pi\theta(-k_0)\delta(k^2-m^2)\;
\nonumber\\
&&\qquad\times
\overline{u}({\imb q}){\cal T}_R(q,k)(\slk+m){\cal T}_A(k,q)
u({\imb q})\; .
\end{eqnarray}

Changing $k\to -k$ in the above formula, and integrating over $k_0$
thanks to the $\delta(k^2-m^2)$ permits to simplify even more the
expression of $\overline{n}$. Noticing that
\begin{eqnarray}
&& {\cal T}_A(-k,q)=-\Big[{\cal T}_R(q,-k)\Big]^*\; ,\\
&& \slk-m=\sum_{\rm spin}v({\imb k})\overline{v}({\imb k})\; ,
\end{eqnarray}
we finally obtain the following compact expression for the average
number of pairs produced per collision at impact parameter ${\imb b}$:
\begin{equation}
\overline{n}=\int{{d^3 {\imb q}}\over{(2\pi)^3 2\omega_{\imb q}}}
\int{{d^3 {\imb p}}\over{(2\pi)^3 2\omega_{\imb p}}}
\Big|
\overline{u}({\imb q}){\cal T}_R(q,-p)v({\imb p})
\Big|^2\; .
\label{eq:nbar}
\end{equation}
Therefore, we have obtained for the average number of produced pairs a
formula similar to Eq.~(\ref{eq:P1}), with however two major
differences: the factor $|\left<0_{\rm out}|0_{\rm in}\right>|^2$ is
not present here, and the formula for $\overline{n}$ involves the
retarded propagator instead of the Feynman propagator.

\subsection{Discussion of earlier literature}
\label{sec:old_lit}
Equations~(\ref{eq:P1}) and (\ref{eq:nbar}) show clearly the main
difference between the approach based on the Dirac equation and the
perturbative approach concerning the ``pair production
cross-section''. Indeed, what is called ``cross-section'' in
\cite{SegevW1,SegevW2,BaltzM1,EichmRSG1}:
\begin{equation}
\sigma_T\equiv \int d^2{\imb b}\; \overline{n}\; ,
\label{eq:sigma_T}
\end{equation}
is the inclusive cross-section\footnote{Integrating over the entire
  range of impact parameters might include contributions from a region
  where our model is not expected to be valid. Cutoffs are implicitly
  understood in this integral, in order to make it finite.} of pair
production (obtained by counting all the pairs produced); while
according to the nature of the diagrams considered in
\cite{IvanoSS1,IvanoSS2}, the object they called ``cross-section'' is
in fact\footnote{Up to the factor $|\left<0_{\rm out}|0_{\rm
      in}\right>|^2$, which they seem to have overlooked.}
\begin{equation}
\sigma_1\equiv \int d^2{\imb b}\; P_1\; ,
\label{eq:sigma_1}
\end{equation}
i.e. the exclusive cross-section to produce exactly one pair (measured
by counting only those events that produce exactly one pair). It is
now obvious that one should not expect an agreement between the two
approaches, since they are calculating different observables.
In fact, it is obvious that $\sigma_1 < \sigma_T$ if there are
collisions that can produce more than one pair. We are now in a
position to discuss in detail the existing literature concerning the
discrepancy between the two approaches.

\subsubsection*{Papers \cite{SegevW1,SegevW2,BaltzM1}}
It seems that those papers overlooked the results of
\cite{ReinhMGS1,RumriMSGG1,WellsOUBS1}. The latter papers, as well as
the above derivation, only connect squares of retarded amplitudes to
average numbers of particles (or other moments of the multiplicity
distribution), but cannot give the pair production {\sl amplitude} in
terms of retarded solutions of the Dirac equation.  The present
derivation shows that \cite{SegevW1,SegevW2,BaltzM1} are not correct
when they call $\overline{u}({\imb q}){\cal T}_R(q,-p)v({\imb p})$ the
pair production amplitude. The only use of this object is via its
square in Eq.~(\ref{eq:nbar}), where it leads to $\overline{n}$.

\subsubsection*{Papers \cite{EichmRSG1,EichmRG1}}
Eichmann and collaborators suggested in \cite{EichmRG1} that the
discrepancy between the two approaches was a consequence of the fact
that crossing symmetry is not valid in the ultra-relativistic limit.
Here is how their argument goes: in \cite{EichmRSG1}, they claim that
solving the Dirac equation in the background field of the two nuclei
leads to the {\sl exact scattering amplitude} of a lepton by the
nuclei. From there, one would have to use crossing symmetry (i.e.
change $p\to -p$ for the incoming electron) in order to obtain the
pair production amplitude, a procedure which they later claimed to be
incorrect because crossing symmetry does not work when the nuclei are
ultra-relativistic \cite{EichmRG1}.

However, the premises of this explanation are incorrect, because
solving the Dirac equation with retarded boundary conditions does not
give the exact scattering amplitude in a relativistic theory if the
background potential is time-dependent. Indeed, in a relativistic
field theory the free retarded and Feynman propagators differ by their
$i\epsilon$ prescription:
\begin{equation}
G_R^0(p)=i{{\slp+m}\over{p^2-m^2+ip_0\epsilon}}\;,\qquad 
G_F^0(p)=i{{\slp+m}\over{p^2-m^2+i\epsilon}}\; .
\end{equation}
The two propagators will lead to equivalent results only if the
background potential cannot change the sign of the energy $p_0$ (the
energy of the incoming lepton is of course positive), i.e. only if the
background potential is static. But if the background potential has
some time dependence, it can change the sign of the energy carried by
the propagator, and the retarded and Feynman propagators will lead to
different results. Basic requirements of any field theory (Lorentz
covariance, unitarity and causality) imply (see \cite{BogolS1}, pages
197-220) that scattering amplitudes are given by the time-ordered
propagator, and not by the retarded one. Physically, the Feynman
propagator takes into account the fact that an external field may
create or annihilate pairs of leptons~\cite{Feynm1}, an effect which
affects lepton scattering. This effect is not taken into account if
one uses the retarded propagator\footnote{See also the section
  \ref{sec:watson} of the present paper, which highlights the
  differences between the retarded and Feynman propagators.}. Of
course, if the external field is static, or if the problem is treated
non-relativistically\footnote{\label{foot:non-relat} The reason why
  scattering in non-relativistic quantum theory can be studied by
  solving the wave equation (even if the potential is non-static) is
  due to the fact that the Feynman and retarded free propagators are
  the same in a non-relativistic theory:
\begin{equation}
G_R^0{}^{\rm n.r.}(p)=G_F^0{}^{\rm n.r.}(p)={{i}\over{p_0-{\imb
p}^2/2m+i\epsilon}}\; ,
\end{equation}
because the propagator associated to Schr\"odinger's equation has a
single pole, at a positive energy.}, pair creation/annihilation
is not possible, and both retarded and Feynman propagators give the
same scattering amplitude.

The authors of \cite{EichmRSG1} also proposed an alternate proof in
section III for their expression of the scattering amplitude, this
time by a direct calculation of Feynman diagrams.  However, the
diagrams displayed in the figure 3 of reference \cite{EichmRSG1} do
not vanish in general if calculated with Feynman propagators.
Although it happens that the diagram with the configuration $ABA$ (see
reference \cite{EichmRSG1} for the notations) of external potentials
vanishes, the diagram $ABBA$ for instance does not.

In conclusion, solving exactly the Dirac equation does not give the
exact lepton scattering amplitude, which makes the discussion of
crossing symmetry irrelevant for the present problem.  Indeed, solving
the Dirac equation (or, equivalently, calculating the retarded
propagator), gives directly $\overline{n}$ thanks to
Eq.~(\ref{eq:nbar}). Of course, since one needs ${\cal T}_R(q,-p)$ and
not ${\cal T}_R(q,p)$, one should be careful not to assume that the
incoming energy is positive when calculating ${\cal T}_R$.  In the
next two sections of the present paper, we discuss in more detail the
differences between the retarded and Feynman propagators in the
background field of the two nuclei. In particular, a closed form
expression for $G_R(q,p)$ is derived (which is valid for energies of
any sign, and therefore can be used in Eq.~(\ref{eq:nbar})). On the
other hand, we show that it is not possible to obtain such a simple
expression for the Feynman propagator.

\subsubsection*{Paper \cite{LeeM1}}
In their paper \cite{LeeM1}, Lee and Milstein proposed a different
explanation for the difference between the two approaches. Starting
from a formula equivalent to our Eq.~(\ref{eq:sigma_T}), and
manipulating possibly ill-defined integrals with special care, they
recover the formula obtained in the references
\cite{IvanoSS1,IvanoSS2} from Feynman diagrams at lowest order in
$Z_1\alpha$. In our language, their identity reads:
\begin{equation}
\int_{{\imb b,p,q}}\Big|\overline{u}({\imb q}){\cal T}_F(q,-p)v({\imb p})\Big|_{Z_1\alpha\ll 1}^2=
\int_{{\imb b,p,q}}\Big|\overline{u}({\imb q}){\cal T}_R(q,-p)v({\imb p})\Big|_{Z_1\alpha\ll 1}^2
\label{eq:milstein}
\end{equation}

However, one should refrain from trying to give a general
interpretation to this result. Indeed, this identification works only
at lowest order in $Z_1$. However, the integrand in the
Eq.~(\ref{eq:sigma_T}) for $\sigma_T$ is known to all orders in both
$Z_1$ and $Z_2$, because the Dirac equation can be solved exactly with
retarded boundary conditions. On the contrary, it is impossible to
calculate all the corrections in $Z_1$ in the perturbative calculation
of \cite{IvanoSS1,IvanoSS2}.  We refer the reader to the sections
\ref{sec:watson} and \ref{sec:small_Z1} of the present paper, where we
show that the retarded and time-ordered amplitudes are qualitatively
different when considered to all orders in $Z_1$ and $Z_2$. In
particular, the retarded amplitude has only a finite number of eikonal
phases, while the time-ordered one can contain arbitrarily many of
them. Therefore, it should be clear that the quantities calculated in
\cite{SegevW1,SegevW2,BaltzM1} and in \cite{IvanoSS1,IvanoSS2} cannot
be the same in general.

The reason why Eq.~(\ref{eq:milstein}) works at lowest order in $Z_1$
is the following: if one considers the time-ordered amplitude only at
lowest order in $Z_1$, it simplifies dramatically into an expression
that has at most two eikonal phases. If one focuses on the eikonal
phases in $\overline{u}({\imb q}){\cal T}_F(q,-p)v({\imb p})$ (see
later the Eq.~(\ref{eq:small_Z1}) for ${\cal T}_F$, with
$p^\pm,q^\pm>0$), one can show that (see \cite{Eichm1}, appendix E):
\begin{eqnarray}
&&\overline{u}({\imb q}){\cal T}_F(q,-p)v({\imb p})\Big|_{Z_1\alpha\ll 1}
\nonumber\\
&&\!\!\!\!\!\!\!\!=\int \cdots\left[
\big[e^{-ie\Lambda({\imb x}_\perp)}-1\big]
+\big[e^{+ie\Lambda({\imb y}_\perp)}-1\big]
+\big[e^{-ie\Lambda({\imb x}_\perp)}-1\big]
\big[e^{+ie\Lambda({\imb y}_\perp)}-1\big]\right]\nonumber\\
&&\!\!\!\!\!\!\!\!=\int \cdots e^{+ie\Lambda({\imb y}_\perp)}
\left[e^{-ie\Lambda({\imb x}_\perp)}
-e^{-ie\Lambda({\imb y}_\perp)}\right]\; ,
\label{eq:small_Z1_amplitude}
\end{eqnarray}
where the dots represent factors that are not needed for the argument.
Doing the same thing for $\overline{u}({\imb q}){\cal
T}_R(q,-p)v({\imb p})$ (see Eq.~(\ref{eq:full-ret}), truncated at
lowest order in $Z_1$), one would get:
\begin{equation}
\overline{u}({\imb q}){\cal T}_R(q,-p)v({\imb p})\Big|_{Z_1\alpha\ll 1}
=\int \cdots
 \Big[e^{-ie\Lambda({\imb
x}_\perp)} -e^{-ie\Lambda({\imb y}_\perp)}\Big]\; ,
\end{equation}
where the dots represent exactly the same factors as in
Eq.~(\ref{eq:small_Z1_amplitude}). One can see that the phase factors
in the retarded amplitude differ from those in the time-ordered
amplitude only by a global phase. It happens that this phase drops out
when one is taking the modulus squared of those quantities and
integrating over the momenta of the leptons, which explains
Eq.~(\ref{eq:milstein}). Because Eq.~(\ref{eq:milstein}) relies on
these properties of eikonal phases, it cannot be true for higher
orders in $Z_1$.

Nevertheless, the calculation of reference \cite{LeeM1} is very
interesting, because it indicates that the integration in
Eq.~(\ref{eq:sigma_T}) should be handled with great care. In
particular, it seems that Coulomb corrections to $\sigma_T$ survive
the integration over impact parameter ${\imb b}$ in the model defined
by Eqs.~(\ref{eq:ur-potentials}) and (\ref{eq:Lagrangian}).

In view of the experimental result of \cite{Vanea1,Vanea2,Vanea3}
which observed an almost exact $Z^2$ scaling of the positron yield
(or, equivalently, the number of produced pairs) apparently
incompatible with large Coulomb corrections, one should mention a
possible shortcoming of such a model. As explained in reference
\cite{LeeM1}, the Coulomb corrections in the integral
Eq.~(\ref{eq:sigma_T}) come entirely from the {\sl point} ${\imb
  k}_\perp=0$. Therefore, they survive in $\sigma_T$ only if there is
no cutoff preventing zero momentum transfers (either intrinsic to a
theoretical calculation performed with a finite $\gamma$, or coming
from the experimental setup).

\section{Propagators in the presence of one nucleus}
\label{sec:1-nucleus}
Even if the seeming discrepancy of the two approaches is now
explained, there is still a paradox remaining. The work of
\cite{SegevW1,SegevW2,BaltzM1,EichmRSG1} indicates that $\sigma_T$
(or, equivalently, the retarded propagator) can be calculated exactly
in the case of the collision of two ultra-relativistic nuclei. On the
contrary, the perturbative expansion of \cite{IvanoSS1,IvanoSS2}
leaves little hope that $\sigma_1$ (or the Feynman propagator) can be
calculated exactly. Why are these two propagators, calculated in the
same background field, so different? This is the question we address
in the next two sections, by showing that in the ultra-relativistic
limit there is a simplification that allows to express in closed form
the retarded propagator, but does not help to calculate the Feynman
propagator.

Let us first consider the case of the background field of only one
nucleus of charge $Z$. It happens that for this case, the perturbative
expansion of the scattering matrix can be summed into a closed
expression when the nucleus is ultra-relativistic. This result will
later appear as a building block in the expression of the propagator
in the case of two nuclei.  For a nucleus moving at the speed of light
in the positive $z$ direction, the Fourier transform of the potential
has the generic form $A^\mu(k)=\delta(k^-)v^\mu_+\Lambda({\imb
k}_\perp)$. For a point-like Coulomb interaction, $\Lambda({\imb
k}_\perp)\sim Ze{\imb k_\perp}^{-2}$, but the following discussion
does not depend on a specific form for $\Lambda({\imb
k}_\perp)$. Therefore, one could possibly take into account effects
like an electromagnetic form factor for the nuclei.

Instead of the propagator itself, it is simpler to deal with the
scattering matrix $T(q,p)$ (we reserve the calligraphic letter ${\cal
  T}$ for scattering matrices in the presence of two nuclei) defined
by
\begin{equation}
G(q,p)\equiv (2\pi)^4\delta(p-q)G^0(p)+G^0(q)T(q,p)G^0(p)\; .
\label{eq:G-T}
\end{equation}
The term of order $n$ in the perturbative expansion of this object is
given by
\begin{eqnarray}
&&\!\!\!\!\!\!\!\!\!\!T_{n}(q,p)=(-ie)^n \int {{d^4k_1}\over{(2\pi)^4}}2\pi\delta(k_1^-)
\Lambda({\imb k}_{1\perp})\cdots
\int{{d^4k_n}\over{(2\pi)^4}}2\pi\delta(k_n^-)\Lambda({\imb k}_{n\perp})
\nonumber\\
&&\!\!\!\!\!\!\!\!\times \slv_+ G^0(p+k_1)\slv_+\cdots 
G^0(p+k_1+\cdots+k_{n-1})\slv_+ (2\pi)^4\delta(p+k_1+\cdots+k_n-q)\; .\nonumber\\
&&
\end{eqnarray}
At this stage, one can use the $\delta(k^-_i)$ to perform for free all
the integrations over the $k^-_i$ components. Using $\slv_+
(\slp+m)\slv_+=2p^- \slv_+$, we obtain
\begin{eqnarray}
&&\!\!\!\!T_{n}(q,p)=(-ie)^n 2\pi\delta(p^--q^-)\slv_+\nonumber\\
&&\quad\!\!\!\!\times
\int\! {{d^2{\imb k}_{1\perp}}\over{(2\pi)^2}}\Lambda({\imb k}_{1\perp})\cdots
\!\int\! {{d^2{\imb k}_{n\perp}}\over{(2\pi)^2}}\Lambda({\imb k}_{n\perp})
(2\pi)^2\delta({\imb p}_\perp+{\imb k}_{1\perp}+\cdots+{\imb k}_{n\perp}-{\imb q}_\perp)\nonumber\\
&&\quad\!\!\!\!\times \int {{dk_1^+}\over{2\pi}}\cdots\int{{dk_n^+}\over{2\pi}} 
2\pi\delta(p^++k_1^++\cdots+k_n^+-q^+)\nonumber\\
&&\quad\!\!\!\!\times
{i\over{p^++k_1^+-{{{\imb \omega}_1^2}\over{2p^-}}+i\epsilon_p}}\cdots
{i\over{p^++k_1^++\cdots+k_{n-1}^+-{{{\imb \omega}_{n-1}^2}\over{2p^-}}+i\epsilon_p}}\; ,
\label{eq:Tn}
\end{eqnarray}
where $\epsilon_p$ is $\epsilon>0$ in the case of the retarded
propagator\footnote{For an advanced propagator, $\epsilon_p$ would be
  $-\epsilon<0$.}, and ${\rm sign}(p^-)\epsilon$ in the case of the
Feynman propagator, and where we denote:
\begin{equation}
\omega_a^2\equiv m^2+({\imb p}_\perp+{\imb k}_{1\perp}+\cdots+{\imb k}_{a\perp})^2\; .
\end{equation}

The next step is to perform the integrals over the $k^+_i$. A
convenient trick is to introduce new variables $A_i\equiv
k_i^++(\omega_{i-1}^2-\omega_i^2)/2p^-$ (and
$\omega_0^2\equiv 2p^+p^-$) and write this integral as
\begin{eqnarray}
&&I_n^+\equiv {1\over{n!}}\!\!\!\!\!\!\sum_{{\rm perms.\ of\ the\ } A_i}
\!\int {{dA_1}\over{2\pi}}\cdots\int{{dA_n}\over{2\pi}} 
2\pi\delta(A_1+\cdots+A_n+{{\omega_n^2}\over{2p^-}}-q^+)\nonumber\\
&& \qquad\qquad\qquad\qquad\qquad\qquad\times{i\over{A_1+i\epsilon_p}}\cdots {i\over{A_1+\cdots+A_{n-1}+i\epsilon_p}}\; .
\end{eqnarray}
 Note that the $1/n!$ is exactly
compensated by the sum over permutations of the $A_i$ (because these
are interchangeable integration variables); it has been inserted for
later convenience. Then, the following combinatoric formula
\begin{equation}
\sum_{\sigma\in {\mathfrak S}_n} {i\over{A_{\sigma(1)}}}\cdots
{i\over{A_{\sigma(1)}+\cdots+A_{\sigma(n-1)}}}=
{i\over{A_1}}\cdots{i\over{A_{n}}}\,{{A_1+\cdots+A_n}\over{i}}\; ,
\end{equation}
where ${\mathfrak S}_n$ is the permutation group of $[1,\cdots,n]$, makes the
various integrations almost independent (they are now coupled only by
the $\delta$ function). At this stage, one begins with the
$\delta$ function to get rid of $A_n$, and then performs
successively the integrals over $A_{n-1}\ldots A_1$ in the complex
plane. The final answer for $I_n^+$ is extremely simple:
\begin{equation}
I_n^+={{({\rm sign}(\epsilon_p))^{n-1}}\over{n!}}\; .
\end{equation}

The transverse integral (second line of Eq.~(\ref{eq:Tn})) factorizes
completely in the space of transverse coordinates, so that we obtain
\begin{equation}
 I^\perp_n=\int d^2{\imb x_\perp}\; [\Lambda({\imb x_\perp})]^n 
e^{i({\imb q_\perp}-{\imb p_\perp})\cdot {\imb x_\perp}}\; ,
\end{equation}
where $\Lambda({\imb x_\perp})$ is the inverse Fourier transform of
$\Lambda({\imb k_\perp})$.

Collecting all the pieces, we have
\begin{equation}
T_{n}(q,p)=2\pi\delta(p^--q^-)\slv_+ {\rm sign}(\epsilon_p)
\!\int\! d^2{\imb x_\perp} {{[-ie\,{\rm sign}(\epsilon_p)
\Lambda({\imb x_\perp})]^n}\over{n!}} 
e^{i({\imb q_\perp}-{\imb p_\perp})\cdot {\imb x_\perp}}\; ,
\end{equation}
and summing over $n$ from $1$ to $+\infty$ to get the full $T(q,p)$ is
now trivial\footnote{For the scattering matrix associated with a
  nucleus moving in the $-z$ direction, replace $p^-,q^-,v_+$ by
  $p^+,q^+,v_-$, and use the appropriate $\Lambda({\imb x}_\perp)$.}:
\begin{equation}
T(q,p)=2\pi\delta(p^--q^-)\slv_+ {\rm sign}(\epsilon_p)
\int d^2{\imb x_\perp} \left[e^{-ie\,{\rm sign}(\epsilon_p)
\Lambda({\imb x_\perp})}-1\right] 
e^{i({\imb q_\perp}-{\imb p_\perp})\cdot {\imb x_\perp}}\; .
\end{equation}

We can now be more specific, and write explicitly the scattering
matrix for the retarded ($\epsilon_p>0$) and for the Feynman (${\rm
  sign}(\epsilon_p)={\rm sign}(p^-)$) propagators\footnote{For the advanced prescription, the result is:
\begin{equation}
\!\!\!\!\!\!\!T_A(q,p)\!=\!-2\pi\delta(p^--q^-)\slv_+\!
\int\!\! d^2{\imb x_\perp}\! \left[e^{ie
\Lambda({\imb x_\perp})}-1\right] 
e^{i({\imb q_\perp}-{\imb p_\perp})\cdot {\imb x_\perp}}
\; .
\label{eq:T-adv}
\end{equation}
}:
\begin{equation}
\!\!\!\!\!\!\!T_R(q,p)\!=\!2\pi\delta(p^--q^-)\slv_+\!
\int\!\! d^2{\imb x_\perp}\! \left[e^{-ie
\Lambda({\imb x_\perp})}-1\right] 
e^{i({\imb q_\perp}-{\imb p_\perp})\cdot {\imb x_\perp}}
\label{eq:T-ret}
\end{equation}
\begin{equation}
T_{F}(q,p)\!=\!
2\pi\delta(p^--q^-)\slv_+ {\rm sign}(p^-)\!
\int \!\!d^2{\imb x_\perp}\! \left[e^{-ie\,{\rm sign}(p^-)
\Lambda({\imb x_\perp})}-1\right] 
e^{i({\imb q_\perp}-{\imb p_\perp})\cdot {\imb x_\perp}}\; .
\label{eq:T-feyn}
\end{equation}
We observe that the retarded and Feynman results differ only by a
${\rm sign}(p^-)$ appearing in two places. This sign will turn out to
be essential when we go to the case of two nuclei, basically because
the interaction with the second nuclei can change the sign of the
$p^-$ of the electron\footnote{One can already notice here that it is
  the retarded version of the scattering which appears in the solution
  of the Dirac solution
  \cite{SegevW1,SegevW2,BaltzM1,EichmRSG1,Baltz2}.  This should not be
  a surprise, since solving the Dirac equation with {\sl initial
    boundary conditions} involves naturally the retarded propagator.
  \cite{EichmRG1} noticed the ${\rm sign}(p^-)$, but attributed to
  some mistake its absence in the solution of the Dirac equation. The
  above considerations show that the solution of the Dirac equation is
  correct, since it is $T_R$ and not $T_F$ that should appear in the
  solution with this type of boundary condition.}.

\begin{figure}[htbp]
\centerline{\resizebox*{4cm}{!}{\includegraphics{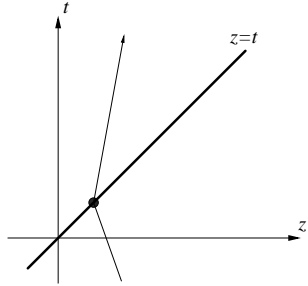}}}
\caption{\sl Schematic representation of the scattering of an $e^-$
  by one ultra-relativistic nucleus. The black dot denotes the
  scattering matrix $T$. The thin lines are free propagators, and the
  thick line is the trajectory of the nucleus.}
\label{fig:1-nucleus}
\end{figure}

Using both Eq.~(\ref{eq:G-T}) and the above results for the scattering
matrix $T$, the interpretation of the result is rather
straightforward: the propagator is the sum of two terms; one of them
is the free propagator $G^0$ (nothing happens to the electron), and
the second term contains the scattering matrix sandwiched between two
free propagators. In space-time, the support of $T$ is on the
hyper-plane $t=z$, where the field of the nucleus is non-zero, and we
can represent the term $G^0 T G^0$ by the diagram of figure
\ref{fig:1-nucleus}.

The formulae of Eqs.~(\ref{eq:T-ret}) and (\ref{eq:T-feyn}) also
illustrate the discussion of section \ref{sec:old_lit}. Indeed, they
show that the retarded and Feynman prescription lead to the same
result in the field of a single nucleus (i.e. in a background field
that can be made static by a change of frame) if $p_0>0$. This is
perfectly consistent with the fact that the difference only shows up
in a time-dependent background field.

\section{Propagators in the presence of two nuclei}
\label{sec:watson}
\subsection{Formal derivation: Watson's series}
Now that we have an exact result in the case of one nucleus, we
present a formal solution for the case of two nuclei that uses the
previously obtained scattering matrices $T(q,p)$ as its building blocks.
This approach allows us to derive some results regarding the retarded
and Feynman propagators in presence of two nuclei, that can be checked
directly from perturbation theory.

Let us assume that we have to solve some generic Lippmann-Schwinger
equation $G=G^0+G^0VG$ (in this section, we use very compact notations
for the sake of brevity; the previous formula is in fact an integral
equation), and that the potential $V$ receives contributions from a
number of different scattering centers: 
$$V=\sum_\alpha V_\alpha\; .$$
Let us assume also that the scattering
matrices $T_\alpha$ for the individual scattering centers are known.
They satisfy
$$T_\alpha=V_\alpha+V_\alpha G^0 T_\alpha\; .$$

Then, the full propagator $G$ resulting from the action of all the
scattering centers can be formally written in terms of the $T_\alpha$.
Indeed, it is a pure matter of algebra to check that the following
object
\begin{equation}
G=G^0+\sum_\alpha G^0 T_\alpha G_\alpha
\end{equation}
is a solution of the full Lippmann-Schwinger equation (see
\cite{GoldbW1}, pages 750-752), provided the $G_\alpha$ satisfy
\begin{equation}
G_\alpha=G^0+\sum_{\beta\not=\alpha}G^0 T_\beta G_\beta\; .
\label{eq:watson}
\end{equation}

This formal solution amounts to a reorganization of the initial
perturbative expansion, which resums infinite subsets of terms
corresponding to the $T_\alpha$.  Despite this achievement, the
problem of finding $G$ in closed form is far from being solved,
because the equations~(\ref{eq:watson}) are coupled integral equations
that are very difficult to solve. The expansion in powers of the
$T_\alpha$ that emerges naturally from them is known in the literature
as Watson's series.

\subsection{The case of moving scattering centers}
\begin{figure}[htbp]
\centerline{\resizebox*{4cm}{!}{\includegraphics{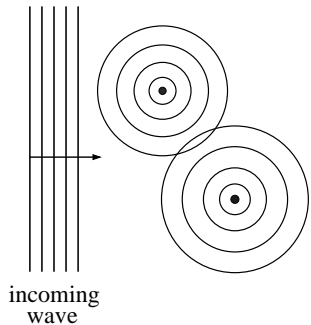}}}
\caption{\sl Illustration of Watson's series in the case of 
  two centers. The scattered waves bounce back and forth between the
  two scattering centers.}
\label{fig:watson}
\end{figure}

The set (\ref{eq:watson}) of coupled equations has a very intuitive
interpretation in the context of wave propagation. Indeed, the
``Lippmann-Schwinger'' equation $\psi=\phi+G^0V\psi$ ($\phi$ being the
incoming wave, $G^0$ being a propagator for the free wave equation)
is solved exactly in the same way by
\begin{eqnarray}
&&\psi=\phi+\sum_\alpha G^0 T_\alpha \psi_\alpha\; ,\nonumber\\
&&\psi_\alpha=\phi+\sum_{\beta\not=\alpha}G^0 T_\beta \psi_\beta\; .
\end{eqnarray}
In this solution, one can interpret $G^0 T_\alpha \psi_\alpha$ as the
partial wave scattered by the scattering center $\alpha$, and
therefore $\psi_\alpha$ is the wave seen by the scatterer $\alpha$. The second
equation then tells that the wave seen by the center $\alpha$ is made
of the initial wave, and of contributions coming from the waves
scattered by all the other centers. This interpretation is illustrated
in figure~\ref{fig:watson} in the case of two centers.
On this classical example, it is also intuitive that multiple
reflections of the wave on the two centers cannot happen if the
centers are moving away at a speed larger than the velocity of wave
propagation.

However, the latter remark is correct only in the case of a
non-relativistic wave equation. In the case of a relativistic wave
equation, the limit where the two centers are moving at the speed of
light is more intricate.  In the next subsection, we show that in the
case of the Dirac equation, a similar conclusion holds for the
retarded propagator, but not for the Feynman propagator.

\subsection{Retarded vs. Feynman propagators}
\subsubsection{Retarded propagator}
\begin{figure}[htbp]
\centerline{\resizebox*{5cm}{!}{\includegraphics{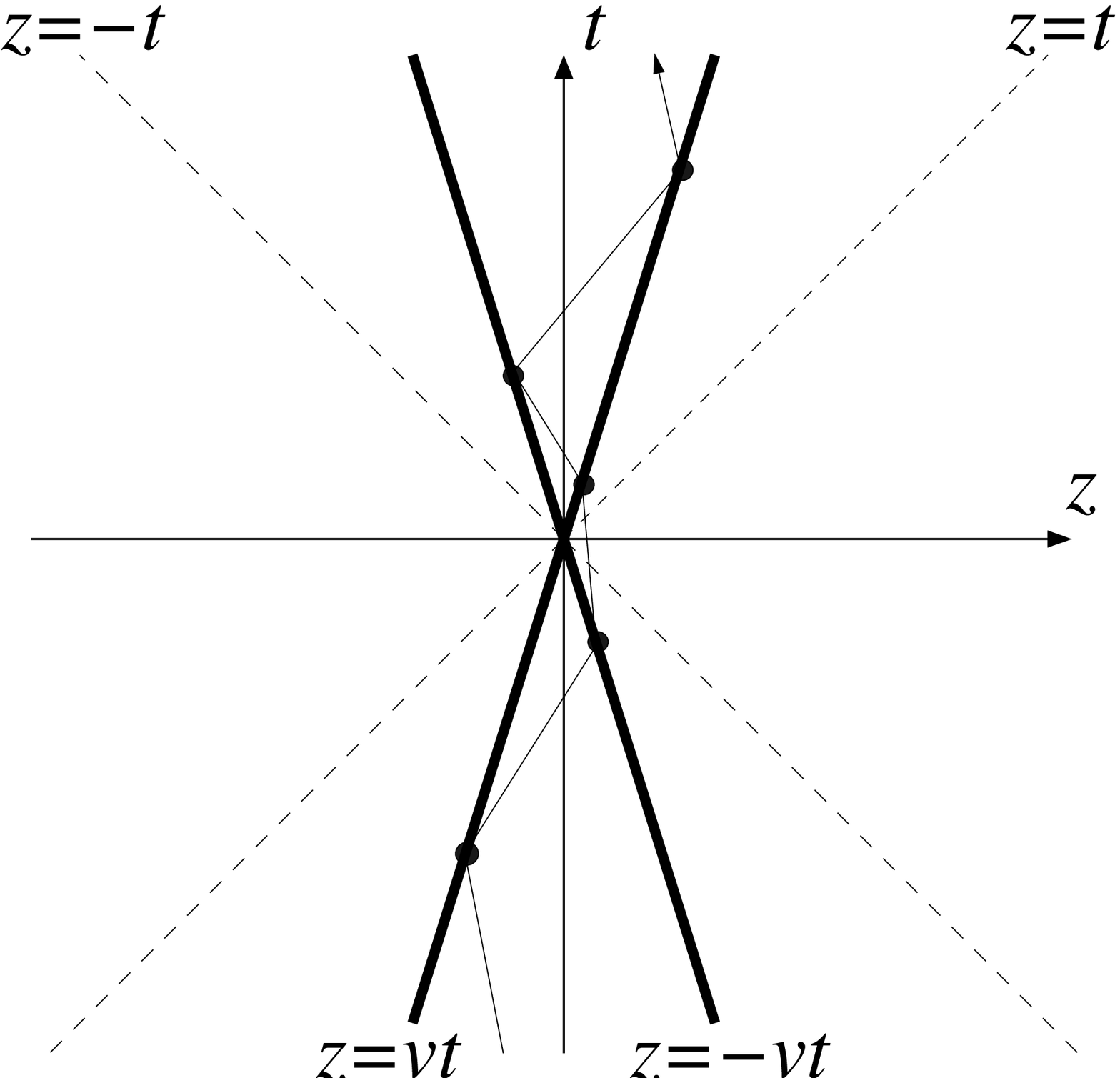}}\hglue 10mm
\resizebox*{5cm}{!}{\includegraphics{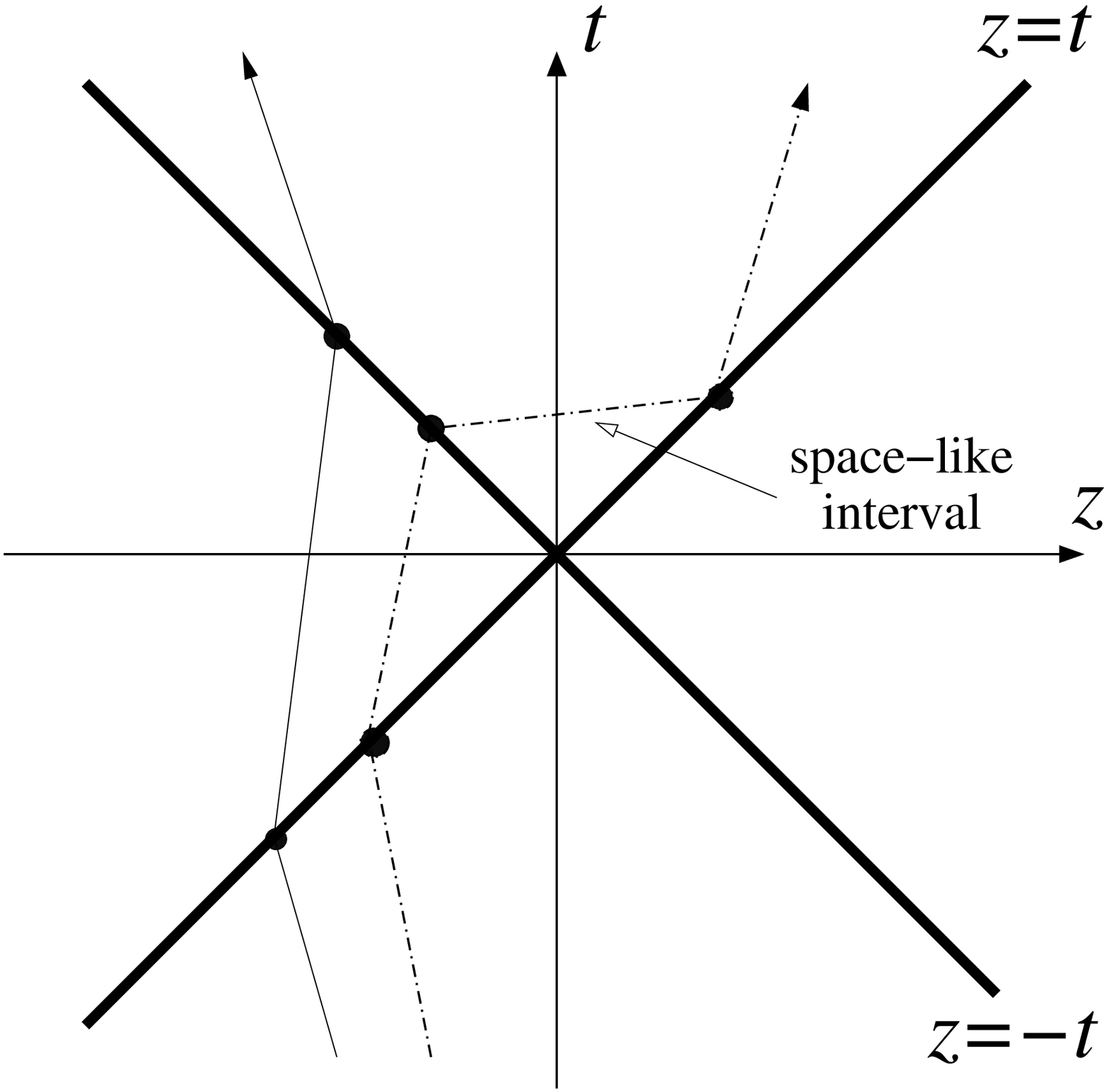}}}
\caption{\sl Typical contribution to the retarded propagator 
  in the case of two nuclei. The thin lines are free retarded
  propagators, and the black dots are scattering matrices $T_R$. Left:
  the two nuclei are non-relativistic (velocity $v<1$). Right:
  ultra-relativistic case; terms with more than two dots are forbidden
  (dashed line).}
\label{fig:ret-prop}
\end{figure}

If $G^0$ is the free retarded propagator, a typical contribution to
the full retarded propagator is represented in the
figure~\ref{fig:ret-prop} (left) when the two nuclei move at a speed
$v$ smaller than the velocity of light. Only this type of contribution
can contribute to the retarded propagator, because the free retarded
propagator connecting each scattering matrix is vanishing outside of
the forward light-cone. This property has also important consequences
in the case where the two nuclei are flying away at the speed of
light, because the separation of two points lying respectively on the
hyper-planes $z=\pm t$ is space-like, except if those points have
$z$'s of the same sign. This restriction forbids terms with more than
two insertions of the scattering matrices (see
figure~\ref{fig:ret-prop}, right).

In the perturbative expansion of the retarded propagator, this
simplification of the ultra-relativistic limit appears in the
following way: terms where interactions with the second nuclei are
inserted between interactions with the first nuclei, like in
\begin{equation}
T_{1R}G_R^0 T_{2R}G_R^0 T_{1R}\; ,
\label{eq:ret-mix}
\end{equation}
are vanishing ($T_{1R}$ and $T_{2R}$ are the retarded scattering
matrices associated to the nuclei $1$ and $2$ respectively, in the
retarded prescription (given by Eq.~(\ref{eq:T-ret}))). Indeed, it is
immediate to check that such a term would have all its poles on the
same side of the real-energy axis. More generally, the only terms
allowed do not alternate interactions with the two nuclei: they can at most
contain one ``packet'' of interactions with one nucleus, followed by a
``packet'' of interactions with the other nucleus. Using the above
symbolic notations (integrations over the momenta exchanged with the
nuclei are implicit), the full retarded propagator in presence of two
ultra-relativistic nuclei reads:
\begin{eqnarray}
G_R=&&G_R^0\nonumber\\
&&
+G_R^0 T_{1R} G_R^0+G_R^0 T_{2R} G_R^0\nonumber\\
&&
+ G_R^0 T_{1R} G_R^0 T_{2R} G_R^0
+G_R^0 T_{2R} G_R^0 T_{1R} G_R^0\; .
\label{eq:full-ret}
\end{eqnarray}
These are the first three orders in the expansion of
Eq.~(\ref{eq:watson}). All the following terms vanish because they
involve factors like Eq.~(\ref{eq:ret-mix}).  This result is precisely
the object that appeared in the solution of the Dirac equation
\cite{SegevW1,SegevW2,BaltzM1,EichmRSG1,Baltz2}, confirming the fact
that this approach in fact derived the retarded propagator. That was
to be expected given the boundary conditions used to solve the Dirac
equation.

\subsubsection{Feynman propagator}
\begin{figure}[htbp]
\centerline{\resizebox*{5cm}{!}{\includegraphics{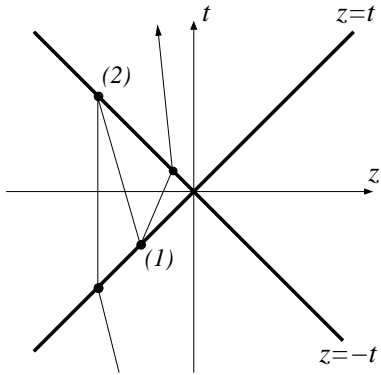}}}
\caption{\sl Example of a contribution to the Feynman propagator 
  that does not vanish in the ultra-relativistic limit. The thin lines
  are free Feynman propagators, and the black dots are scattering
  matrices $T_F$. Point (1): a pair is created; point (2): a pair is
  annihilated.}
\label{fig:feyn-prop}
\end{figure}

Unlike the free retarded propagator, the free Feynman propagator can
connect any pair of points in space-time. Therefore, the
ultra-relativistic limit does not forbid any contribution to the
Feynman propagator. Physically, the additional terms correspond to the
creation of additional $e^+e^-$ pairs, which are annihilated later so
that only one electron is present when $t\to +\infty$. This is
illustrated in figure~\ref{fig:feyn-prop}.

In perturbation theory, the main difference with the retarded case is
that the analogue of Eq.~(\ref{eq:ret-mix}):
\begin{equation}
T_{1F} G_F^0 T_{2F} G_F^0 T_{1F}\; ,
\end{equation}
is not vanishing. This can be traced back into the fact that the free
Feynman propagator has poles on both sides of the real-energy axis,
and this in turn is related to the fact that the free Feynman
propagator can connect points with any time-ordering. As a
consequence, the Watson's series for the Feynman propagator is
infinite even in the ultra-relativistic limit:
\begin{eqnarray}
G_F=&&G_F^0\nonumber\\
&&
+G_F^0 T_{1F} G_F^0+G_F^0 T_{2F} G_F^0\nonumber\\
&&
+ G_F^0 T_{1F} G_F^0 T_{2F} G_F^0
+G_F^0 T_{2F} G_F^0 T_{1F} G_F^0\nonumber\\
&&+ G_F^0 T_{1F} G_F^0 T_{2F} G_F^0 T_{1F} G_F^0
+G_F^0 T_{2F} G_F^0 T_{1F} G_F^0 T_{2F} G_F^0\nonumber\\
&&+\cdots\; .
\label{eq:watson-feynman}
\end{eqnarray}

One can note that the difference between the retarded and Feynman
propagators is a feature specific to the relativistic nature of the
Dirac equation. Indeed, for a non-relativistic wave equation like
Schr\"odinger's equation, the retarded and Feynman $i\epsilon$
prescriptions give the same propagator because the propagator has a
unique pole, which has a positive energy (see footnote
\ref{foot:non-relat}). This remark is in agreement with the fact that
the surviving terms involve the creation/annihilation of additional
pairs, a purely relativistic effect. This interpretation in terms of
pair creation/annihilation also highlights why the retarded and
Feynman propagators are equally simple in the case of one nucleus,
whereas they are not in the case of two nuclei: this is due to the
fact that pairs cannot be produced by a single nucleus.

\section{Unitarity}
\label{sec:unitarity}
\subsection{Calculation of $|\left<0_{\rm out}|0_{\rm
      in}\right>|^2$} So far, we have said nothing about the factor
$\left<0_{\rm out}|0_{\rm in}\right>$ which appeared in the expression
of the pair production amplitude, besides the fact that this factor is
not just a phase. In this paragraph, we derive an expression for this
vacuum-vacuum amplitude, that depends only on the Feynman propagator.
In order to calculate $|\left<0_{\rm out}|0_{\rm in}\right>|^2$, we
start from the well known fact that the corresponding amplitude is the
exponential of the sum of vacuum-vacuum diagrams \cite{ItzykZ1}:
\begin{equation}
\left<0_{\rm out}|0_{\rm in}\right>=e^{iV}\; ,
\label{eq:vacuum-start}
\end{equation}
with
\setbox1=\hbox to 9cm{\resizebox*{9cm}{!}{\includegraphics{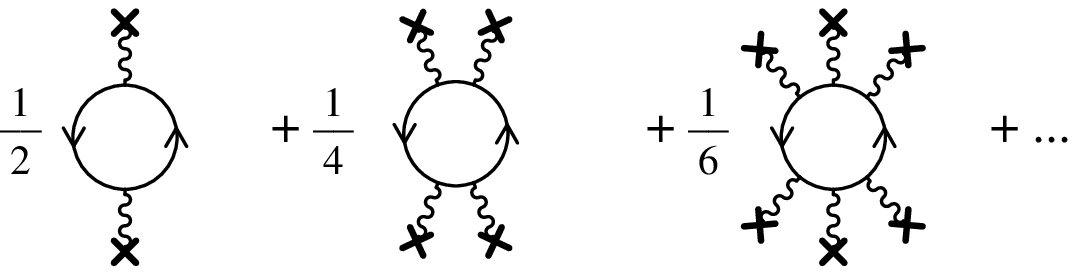}}}
\begin{equation}
  iV\equiv\;\raise -11mm\box1\; ,
\label{eq:vacuum}
\end{equation}
where the coefficients $1/2$, $1/4, \cdots$ are the symmetry factors
of the the corresponding diagrams. Then, the probability of vacuum to
vacuum transition is given by
\begin{equation}
|\left<0_{\rm out}|0_{\rm in}\right>|^2=e^{-2\,{\rm Im}\,V}\; .
\end{equation}
To proceed, one could use cutting rules\footnote{In order to use cutting
  rules, it is essential to notice that, since the sign of the energy
  flowing in the photon lines is not fixed, one cannot exclude
  contributions where the cut divides the diagram in more than two
  connected pieces. This is precisely what happens in terms like the
  second diagram of figure \ref{fig:poisson}.} in order to compute the
imaginary part of the vacuum-vacuum diagrams. However, this approach
makes cumbersome the tracking of symmetry factors. For this reason, it
is much simpler to remark that $V$ is also given by\footnote{This
  formula, together with Eq.~(\ref{eq:vacuum-start}), is well known
  for the vacuum-to-vacuum amplitude. In the literature on pair
  production in nuclear collisions, it already appears in
  \cite{HenckTB2} (see their Eq.~(45)). However, \cite{HenckTB2} does
  not work out its modulus squared $|\left<0_{\rm out}|0_{\rm
      in}\right>|^2$.}
\begin{equation}
iV={\rm Tr}\;\ln (1+ie\,\slA\, G_F^0)\; ,
\label{eq:vacuum-log}
\end{equation}
where the symbol ${\rm Tr}$ denotes a trace on Dirac's indices, as
well as a trace over space-time. It is trivial to check that the
expansion of the logarithm generates the series of diagrams in $V$,
with the correct symmetry factors and a minus sign for the fermion
loop. To begin with, one must write the Lippmann-Schwinger equation
for the full Feynman scattering matrix\footnote{Let us recall that
  $G_F=G_F^0+G_F^0{\cal T}_F G_F^0$.} (in presence of the two nuclei).
Formally, it reads
\begin{equation}
{\cal T}_F=-ie\,\slA\,-ie\,\slA\, G_F^0 {\cal T}_F
=-ie\,\slA\,-{\cal T}_F G_F^0 ie\,\slA\,\; ,
\end{equation}
from which we deduce\footnote{${\cal T}_F^*\equiv \gamma^0 {\cal T}_F^\dagger \gamma^0$, $G_F^0{}^*\equiv \gamma^0 G_F^0{}^\dagger \gamma^0$, $\slA\in {\mathbb R}$.}
\begin{equation}
{\cal T}_F+{\cal T}_F^*=-{\cal T}_F^*(G_F^0+G_F^0{}^*){\cal T}_F
=-{\cal T}_F^*(\rho^{(+)}+\rho^{(-)}){\cal T}_F\; ,
\end{equation}
where we denote $\rho^{(\pm)}(p)\equiv2\pi\theta(\pm
p_0)(\slp+m)\delta(p^2-m^2)$.

Noticing then that closed loops of retarded propagators are zero, and
that $G_R^0=G_F^0-\rho^{(-)}$, we have
\begin{eqnarray}
0&&={\rm Tr}\;\ln(1+ie\,\slA\,G_R^0)\nonumber\\
&&={\rm Tr}\;\ln(1+ie\,\slA\,G_F^0-ie\,\slA\,\rho^{(-)})\nonumber\\
&&={\rm Tr}\;\ln(1+ie\,\slA\,G_F^0+(1+ie\,\slA\,G_F^0){\cal T}_F\rho^{(-)})\nonumber\\
&&=iV+{\rm Tr}\;\ln(1+{\cal T}_F\rho^{(-)})\; .
\end{eqnarray}
From there, one obtains
\begin{eqnarray}
-2{\rm Im}\,V=i(V-V^*)
&&=-{\rm Tr}\; \ln((1+{\cal T}_F^*\rho^{(-)})
(1+{\cal T}_F\rho^{(-)}))\nonumber\\
&&=-{\rm Tr}\;\ln((1
+{\cal T}_F^*\rho^{(-)}{\cal T}_F\rho^{(-)}
+({\cal T}_F+{\cal T}_F^*)\rho^{(-)}))\nonumber\\
&&=-{\rm Tr}\;\ln(1-{\cal T}_F^*\rho^{(+)}
{\cal T}_F\rho^{(-)})\; .
\end{eqnarray}
Therefore, we have
\begin{equation}
|\left<0_{\rm out}|0_{\rm in}\right>|^2=e^{-{\rm Tr}\;\ln(1-{\cal T}_F^*\rho^{(+)}
{\cal T}_F\rho^{(-)})}\; .
\label{eq:vac-vac}
\end{equation}
From this formula, we know $\left<0_{\rm out}|0_{\rm in}\right>$, up
to an irrelevant phase. This expression shows that the vacuum-vacuum
amplitude is completely determined once we have set up some
approximation scheme that gives the Feynman propagator. The very
reason for this is the nature of the Lagrangian of
Eq.~(\ref{eq:Lagrangian}). Indeed, this Lagrangian implies that Wick's
theorem holds\footnote{This would not be true if we had kept the
  photon kinetic term $F_{\mu\nu}F^{\mu\nu}/4$ in the Lagrangian.}, so
that we can express the full $2n$-point Green's functions in terms of
$G_F$ only.

\subsection{Probability to produce ${\imb n}$ pairs}
Since there is some confusion in the literature regarding unitarity
in ultra-relativistic heavy ion collisions
\cite{BertuB1,Eichl1,Baur1,Baur2,EichmRG2,VidovGS1,BestGS1,RhoadW1},
it is useful to derive the probability of production of $n$ pairs.
The corresponding transition amplitude is
\begin{equation}
M_n(\{{\imb p}_i,{\imb q}_i\})\equiv
\left<e^+({\imb p}_1)\cdots e^+({\imb p}_n) 
e^-({\imb q}_1)\cdots e^-({\imb q}_n){}_{\rm out}|0_{\rm in}\right>\; .
\end{equation}
This amplitude can be related by a reduction formula similar to
Eq.~(\ref{eq:reduction}) to the Fourier transform of the amputated
$2n$-point time-ordered correlator (with appropriate spinors for the
final states). Again, we pull out the factor $\left<0_{\rm out}|0_{\rm
    in}\right>$, in order to get rid of vacuum-vacuum diagrams in the
second factor:
\begin{equation}
\left<0_{\rm out}|0_{\rm
    in}\right>{{
\left<0_{\rm out}|
{\rm T} \overline{\psi}(y_1)\cdots \overline{\psi}(y_n)\psi(x_1)\cdots\psi(x_n)
|0_{\rm
    in}\right>
}\over{\left<0_{\rm out}|0_{\rm
    in}\right>}}\equiv\left<0_{\rm out}|0_{\rm
    in}\right> G_{2n}(\{x_i,y_i\}) \; .
\end{equation}
At this stage, it is important to notice that since the Lagrangian
Eq.~(\ref{eq:Lagrangian}) does not contain any dynamical field that
couples to the fermions, Wick's theorem applies to the second factor
$G_{2n}(\{x_i,y_i\})$ so that we can write it in terms of the full
Feynman propagator:
\begin{equation}
G_{2n}(\{x_i,y_i\})=\sum_{\sigma\in {\mathfrak S}_n}\epsilon(\sigma)
G_F(x_1,y_{\sigma(1)})\cdots G_F(x_n,y_{\sigma(n)})\; ,
\end{equation}
where $\epsilon(\sigma)$ is the signature of the permutation $\sigma$
(required when permuting fermion fields). This property extends to the
amputated correlators and their Fourier transform, so that we can
write directly the transition amplitude as\footnote{This formula is
  equivalent to the Eq.~(11) of \cite{HenckTB2}. However, these
  authors do not make use of it to calculate the exact $P_n$ (see our
  Eq.~(\ref{eq:Pn})).}
\begin{eqnarray}
&&M_n(\{{\imb p}_i,{\imb q}_i\})=\left<0_{\rm out}|0_{\rm in}\right>
\sum_{\sigma\in{\mathfrak S}_n}\epsilon(\sigma)
[\overline{u}({\imb q}_{\sigma(1)}){\cal T}_F(q_{\sigma(1)},-p_1) v({\imb p}_1)]
\cdots\nonumber\\
&&\qquad\qquad\qquad\qquad\qquad\qquad\qquad\cdots
[\overline{u}({\imb q}_{\sigma(n)}){\cal T}_F(q_{\sigma(n)},-p_n) v({\imb p}_n)]\; ,
\end{eqnarray}
where ${\cal T}_F$ is the Feynman scattering matrix in presence of the
two nuclei.

The integrated probability of producing exactly $n$ pairs is then (we
have changed $p_i\to -p_i$ in the second line)
\begin{eqnarray}
&&P_n\equiv {1\over{n!^2}}\prod_{i=1}^{n}
\int{{d^4p_i}\over{(2\pi)^4}}{{d^4q_i}\over{(2\pi)^4}}
\widetilde{\rho}^{(+)}(p_i)\widetilde{\rho}^{(+)}(q_i)
\left|M_n(\{{\imb p}_i,{\imb q}_i\})\right|^2
\nonumber\\
&&={{|\left<0_{\rm out}|0_{\rm in}\right>|^2}\over{n!}}
\prod_{i=1}^{n}
\int{{d^4p_i}\over{(2\pi)^4}}{{d^4q_i}\over{(2\pi)^4}}
\widetilde{\rho}^{(-)}(p_i)\widetilde{\rho}^{(+)}(q_i)\nonumber\\
&&\,\times
\sum_{\sigma\in{\mathfrak S}_n}\epsilon(\sigma)(-1)^n\prod_{i=1}^{n}
\left[\overline{u}({\imb q}_i){\cal T}_F(q_i,p_i)(\slp_i+m)
{\cal T}_F^*(p_i,q_{\sigma(i)})u({\imb q}_{\sigma(i)})
\right]\; ,
\end{eqnarray}
where $\widetilde{\rho}^{(\pm)}(p)\equiv 2\pi\theta(\pm
p_0)\delta(p^2-m^2)$.  We can see now that we obtain closed chains
like:
\begin{equation}
{\rm tr}\,[{\cal T}_F(q_1,p_1)(\slp_1+m){\cal T}_F^*(p_1,q_2)(\slq_2+m){\cal T}_F(q_2,p_2)
(\slp_2+m){\cal T}_F^*(p_2,q_1)(\slq_1+m)]\; .
\end{equation}
For instance, for $n=2$, cycles come in two sizes, illustrated on
figure \ref{fig:poisson}. A systematic tool to construct these loops
is the (unique) decomposition of permutations in products of disjoint
circular permutations. It is a trivial matter of combinatorics to find
that the number of $p$-cycles in ${\mathfrak S}_n$ is $n!/p(n-p)!$,
and that the number of permutations that are made of $a_1$ $1$-cycles,
$a_2$ $2$-cycles,$\ldots$,$a_n$ $n$-cycles ($a_1+2a_2+\cdots +na_n=n$)
is:
\begin{equation}
{{n!}\over{a_1!\cdots a_n!}}\; {1\over{1^{a_1}}}\cdots {1\over{n^{a_n}}}\; .
\end{equation}
The signature of such a permutation is $\epsilon(\sigma)=(-1)^n
\prod_i (-1)^{a_i}$. With the following compact notations for
``links''\footnote{Note that $L=0$ in a background field that cannot
  produce pairs, like the field of a single nucleus for instance.} and
``loops'':
\begin{eqnarray}
&&L(p^\prime,p)\equiv 
\int{{d^4q}\over{(2\pi)^4}}  
{\cal T}_F^*(p^\prime,q)\rho^{(+)}(q){\cal T}_F(q,p)\rho^{(-)}(p)\; ,\nonumber\\
&&{\rm Tr}L^n\equiv \int \left[\prod_{i=1}^n {{d^4p_i}\over{(2\pi)^4}}\right]
{\rm tr}[L(p_1,p_2)L(p_2,p_3)\cdots L(p_{n},p_1)]\; ,
\label{eq:L}
\end{eqnarray}
the total probability to produce $n$ pairs is
\begin{equation}
P_n=|\left<0_{\rm out}|0_{\rm in}\right>|^2 \sum_{a_1+2a_2+\cdots +na_n=n}\;
\prod_{i=1}^n \left[{{(-1)^{a_i}}\over{a_i!}}\;
\left({{{\rm Tr}L^i}\over{i}}\right)^{a_i}\right]\; .
\label{eq:Pn}
\end{equation}
We should emphasize here the fact that the individual $P_n$ are
functions of the Feynman scattering matrix ${\cal T}_F$, while the
average number of pairs $\overline{n}=\sum_n nP_n$ has a simple
expression in terms of the retarded scattering matrix ${\cal T}_R$
(see Eq.~(\ref{eq:nbar})).

\subsection{Another look at unitarity}
We can now check unitarity without making any approximation, by using
the previous formula for the $P_n$, and Eq.~(\ref{eq:vac-vac}) for the
vacuum-vacuum transition probability:
\begin{eqnarray}
\sum_{n=0}^{+\infty}P_n&&=|\left<0_{\rm out}|0_{\rm in}\right>|^{2}
\sum_{n=0}^{+\infty}\; \sum_{a_1+2a_2+\cdots +na_n=n}\;
\prod_{i=1}^n \left[{{(-1)^{a_i}}\over{a_i!}}\;
\left({{{\rm Tr}L^i}\over{i}}\right)^{a_i}\right]\nonumber\\
&&=e^{-{\rm Tr}\;\ln(1-{\cal T}_F^*\rho^{(+)}
{\cal T}_F\rho^{(-)})}e^{{\rm Tr}\;\ln(1-L)}=1\; .
\label{eq:unitarity}
\end{eqnarray}
The compensation of the two exponentials is obvious from the first of
Eqs.~(\ref{eq:L}). 

It therefore clarifies the long standing ``unitarity problem''
floating around in the literature related to pair production in
nuclear collisions
\cite{BertuB1,Eichl1,Baur1,Baur2,EichmRG2,VidovGS1,BestGS1,RhoadW1}.
The problem can be stated as follows: perturbation theory seems to
lead to cross sections that are too large in order to comply with
unitarity (or to production probabilities larger than $1$).  It
appears that the factor $\left<0_{\rm out}|0_{\rm in}\right>$ (the
modulus of which is smaller than one) has been overlooked\footnote{Two
  exceptions are \cite{HenckTB2} and \cite{BestGS1}, which noticed
  that unitarity is related to the vacuum to vacuum amplitude.
  However, the authors of \cite{BestGS1} calculated this amplitude by
  requiring that unitarity is preserved (in addition, they only did
  that in an approximation that leads to a Poissonian distribution).
  One needs an independent calculation of $\left<0_{\rm out}|0_{\rm
      in}\right>$ in order to claim that this factor restores
  unitarity. The authors of \cite{HenckTB2} give an exact expression
  of $\left<0_{\rm out}|0_{\rm in}\right>$ equivalent to our
  Eq.~(\ref{eq:vacuum-log}), but unfortunately do not exploit it to check
  unitarity.} in the literature: only the connected piece of Feynman
diagrams has been considered, but not the disconnected vacuum-vacuum
diagrams. In this paper, we have shown that the factor $\left<0_{\rm
    out}|0_{\rm in}\right>$ naturally emerges from the reduction
formula for pair production (and therefore that calculating only the
connected diagrams leads to an incomplete result), and that this
factor restores unitarity.

Moreover, since the factor $\left<0_{\rm out}|0_{\rm in}\right>$
depends only on the Feynman propagator, the above considerations
provide a way to make approximations that preserve unitarity. It is
sufficient to use the same approximate $G_F$ in the calculation of the
$P_n$ and in the calculation of the vacuum-vacuum amplitude.

For such an approximation scheme to be consistent, we have also to
verify that it gives positive $P_n$.  Indeed, the fact that the sum of
Eq.~(\ref{eq:unitarity}) is $1$ leaves open the possibility that some
probabilities could come out negative.  To prove that all the
probabilities are positive (and hence smaller than one, because their
sum is one), it is convenient to introduce a ``generating function''
for the probabilities $P_n$:
\begin{equation}
F(x)\equiv e^{-{\rm Tr}\,\ln(1-L)}e^{{\rm Tr}\,\ln(1-xL)}\; ,
\end{equation}
such that $F(1)=1$ and $P_n=F^{(n)}(0)/n!$. Then, one notices that it
can be rewritten as
\begin{equation}
F(x)= e^{-{\rm Tr}\,\ln(1+t t^\dagger)}e^{{\rm Tr}\,\ln(1+x t t^\dagger)}\; ,
\end{equation}
with $t(q,p)\equiv \overline{u}(q) {\cal T}_F(q,p) v(p)$. The operator
$t t^\dagger$ is positive, and we can write the generating function in
terms of its eigenvalues $\tau_i$ , which are positive:
\begin{equation}
F(x)=\prod_i {{1+\tau_i x}\over{1+\tau_i}}\; .
\end{equation}
We therefore see that all the coefficients in the Taylor expansion of
$F(x)$ are positive independently of the approximation made for ${\cal
T}_F$, which proves that all the $P_n$ are positive in this framework.

\subsection{Nature of the multiplicity distribution}
One can also note that Eq.~(\ref{eq:Pn}) for the probability of
producing $n$ pairs is not a Poisson distribution. This fact
contradicts the solution proposed in \cite{Baur2,RhoadW1} for the
unitarity problem, where a Poissonian distribution is obtained.
However, one can see from Eq.~(\ref{eq:Pn}) the nature of the
approximation that would lead to such a distribution: a Poisson
distribution is obtained if one drops all the ${\rm Tr}L^n$ for $n>1$.
Indeed, this drastic (and a priori not justified) simplification leads
to
\begin{equation}
P_n\to e^{-(-{\rm Tr}L)} {{(-{\rm Tr}L)^n}\over{n!}}\; ,
\label{eq:Pn-approx}
\end{equation}
a Poisson distribution of average $-{\rm Tr}L$. This explains the
somewhat confusing statement of \cite{Baur2,RhoadW1,HenckTB3,HenckTB1}
saying that the ``perturbation theory prediction'' for the probability
to produce one pair (i.e. $P_1$ in which one would forget the factor
$|\left<0_{\rm in}|0_{\rm out}\right>|^2$, that is $-{\rm Tr}L$ in our
notations) should not be interpreted as a probability (because it can
be larger than $1$) but instead should be reinterpreted as the average
number of pairs produced in a nuclear collision. Modulo the
approximation of Eq.~(\ref{eq:Pn-approx}), we agree with this
statement, except for one thing: $-{\rm Tr}L$ is not the perturbation
theory prediction for the probability to produce one pair;
perturbation theory applied correctly indicates that this probability
is $-|\left<0_{\rm out}|0_{\rm in}\right>|^2\,{\rm Tr}L < 1$. In fact,
the present analysis makes the following clear: even if ${\cal T}_F$
(and hence $L$) were known exactly, using $-{\rm Tr}L$ for the
probability to produce one pair would still violate unitarity.  The
unitarity problem does not really come from perturbation theory, but
from forgetting the contribution of the vacuum-vacuum amplitude.

Thanks to the generating function introduced above, it is easy to find
what the average number of produced pairs is\footnote{One can check
  that this formula is equivalent to Eq.~(\ref{eq:nbar}), by using the
  Lippmann-Schwinger equations for ${\cal T}_F$ and ${\cal T}_R$. One
  can obtain successively:
  \begin{eqnarray}
    &&{\cal T}_F=[1-{\cal T}_R\rho^{(-)}]^{-1}{\cal T}_R\; ,\nonumber\\
    &&L=[1-{\cal T}_R^*\rho^{(-)}]^{-1}[{\cal T}_R^*\rho^{(+)}{\cal T}_R\rho^{(-)}][1-{\cal T}_R\rho^{(-)}]^{-1}\; ,\nonumber\\
    &&[1-{\cal T}_R^*\rho^{(-)}][1-{\cal T}_R\rho^{(-)}]-[{\cal T}_R^*\rho^{(+)}{\cal T}_R\rho^{(-)}]=1\; ,
  \end{eqnarray}
    and finally
    \begin{equation}
    \overline{n}=-{\rm Tr}\;[L(1-L)^{-1}]=-{\rm Tr}\,[{\cal T}_R^*\rho^{(+)}{\cal T}_R\rho^{(-)}]\; ,
    \end{equation}
    which can be cast into Eq.~(\ref{eq:nbar}) thanks to
    $\slq+m=\sum_{\rm spin}u({\imb q})\overline{u}({\imb q})$ and
    $\slp-m=\sum_{\rm spin}v({\imb p})\overline{v}({\imb
      p})$.}:
\begin{equation}
\overline{n}\equiv\sum_{n=0}^{+\infty} nP_n=F^\prime(1)=-{\rm Tr}[L(1-L)^{-1}]
=-\sum_{n=1}^{+\infty}{\rm Tr}L^n\; .
\end{equation}
By the same method, we can find the variance of the number of pairs:
\begin{equation}
\overline{n^2}-\overline{n}^2=-{\rm Tr}[L(1-L)^{-1}]-{\rm Tr}[L^2(1-L)^{-2}]\; .
\end{equation}
\begin{figure}[htbp]
\centerline{\resizebox*{11cm}{!}{\includegraphics{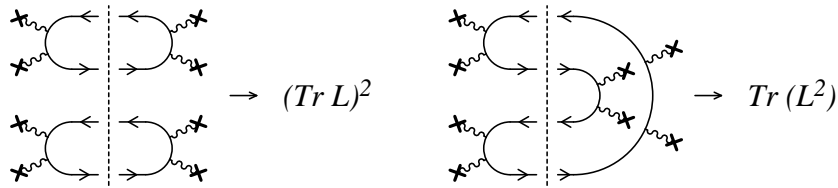}}}
\caption{\sl Two types of pairings in the probability to produce 
  two pairs. Even if the amplitude factorizes because we neglect
  photons connecting directly the various fermion lines, the
  cross-section does not.}
\label{fig:poisson}
\end{figure}
Since $\overline{n}\not=\overline{n^2}-\overline{n}^2$, the exact
probability distribution cannot be a Poisson distribution. This is a
consequence of the permutations in the final state when squaring the
amplitude, which can lead to assign to the same pair\footnote{The
problem comes from the fact that particles are produced in pairs
here. As an example of model in which particles are produced
individually, one can consider a toy model in which a scalar field
$\phi(x)$ is coupled to a background classical source $j(x)$ by ${\cal
L}_{\rm int}\equiv j(x)\phi(x)$. This model is exactly solvable, and
the production probability is found to be Poissonian (see
\cite{ItzykZ1}, pages 163-170).  It is also possible to understand
from this remark why \cite{Baur2} obtained a Poisson
distribution. Indeed, this paper modeled $e^+ e^-$ pairs as {\sl
elementary} fields (quasi-bosons), so that the problem of pairings in
the final state never shows up.} the electron of one fermion line and
the positron from another fermion line. This is illustrated in figure
\ref{fig:poisson} for the case of two pairs, where one can see clearly
the origin of the terms $({\rm Tr}\,L)^2$ and ${\rm Tr}\,L^2$. One
also sees that the term in ${\rm Tr}\,L^2$ correlates the emission of
the two pairs, and therefore prevents the probability distribution
from being Poissonian\footnote{Another argument against a Poisson
distribution is that it would contradict the fact that $\overline{n}$
can be calculated exactly, while the individual $P_n$ cannot.}. This
explains the fact that a Poisson distribution is obtained only if one
neglects the ${\rm Tr}\,L^n$ for $n>1$.

The authors of \cite{HenckTB2} attempted to size the departure from a
Poisson distribution in the Magnus model, and found it to be around
$1\%$, which may justify for most practical purposes to calculate only
$\overline{n}$ (which can be calculated exactly according to our
analysis) and plug it in a Poisson formula for $P_n$. This is the
approach followed in \cite{HenckTB3,HenckTB1}. \cite{HenckTB3} started
from the lowest order (only one photon is exchanged with each nucleus,
and $\left<0_{\rm out}|0_{\rm in}\right>$ is approximated by 1)
$P_1$ in the background field of two nuclei, ``reinterpreted'' as
$\overline{n}$.  \cite{HenckTB1} followed the same approach by
starting from the formula given in Eq.~(53) of \cite{BaltzM1}, which
was assumed to be the exact formula for the pair production
amplitude. If this assumption were correct, its square should be the
exact probability to produce one pair, and should therefore be smaller
than $1$. This is not the case as noted in \cite{HenckTB1}, who
``reinterpreted'' this quantity as the average number of pairs
produced in a collision in order to save unitarity. These authors did
not realize that their remark was directly pointing to the result we
have proven in our section \ref{sec:reduction}: that the retarded
solution of the Dirac equation does not give the pair production
probability, but rather the average number of pairs (without any need
to ``reinterpret'' anything).

\section{Strategies of approximation for $G_F(x,y)$}
\label{sec:approx}
\subsection{Limit $Z_1 \alpha \ll 1$}
\label{sec:small_Z1}
Since the series in Eq.~(\ref{eq:watson-feynman}) giving the full
Feynman propagator cannot be summed exactly, one must use
approximations in order to simplify it. We present in this section two
different approximation schemes which lead to closed form expressions
for the Feynman propagator.

\begin{figure}[htbp]
\centerline{\resizebox*{5cm}{!}{\includegraphics{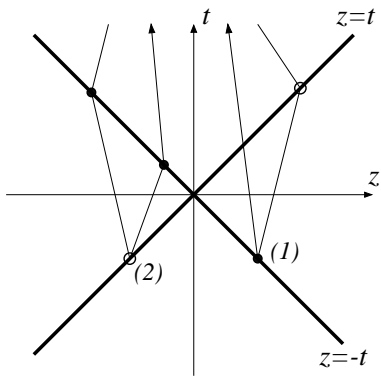}}}
\caption{\sl Dominant pair production mechanisms in the approximation
  $Z_1\alpha\ll 1$. The thin lines are free Feynman propagators, the black
  dots are scattering matrices, and the open dots denote interactions
  at lowest order.}
\label{fig:approx-small-Z1}
\end{figure}

The simplest approximations one can think of are truncations of the
Watson series. Such a simplification is obtained in a natural way if
one assumes that only one nucleus has a large electric charge. If
$Z_1$ is small enough so that we have
\begin{equation}
Z_1\alpha \ll 1 \sim Z_2 \alpha\; ,
\label{eq:condZ}
\end{equation}
then we can neglect most of the corrections in $Z_1\alpha$. Since we
are interested in pair production, the kinematics requires at least one
interaction with the first nucleus (otherwise, the pair cannot be
produced on-shell). We are therefore going to keep only terms with at
most one insertion of $\,\slA_1$. In that approximation, $T_{1F}$ is
just $-ie\slA_1$, and we drop any term with two or more insertions of
$T_{1F}$ in the Watson series. It is immediate to verify that the only
terms left in $G_F$ are
\begin{eqnarray}
G_F\approx&&G_F^0\nonumber\\
&&
+G_F^0 (-ie\slA_1) G_F^0+G_F^0 T_{2F} G_F^0\nonumber\\
&&
+ G_F^0 (-ie\slA_1) G_F^0 T_{2F} G_F^0
+G_F^0 T_{2F} G_F^0 (-ie\slA_1) G_F^0\nonumber\\
&&+G_F^0 T_{2F} G_F^0 (-ie\slA_1) G_F^0 T_{2F} G_F^0\; .
\label{eq:small_Z1}
\end{eqnarray}
This approximation for $G_F$ is in fact the starting point used by
\cite{IvanoM1,IvanoSS1,IvanoSS2} in their approach to the problem of
pair production. The terms that participate to pair production (i.e.
having at least one interaction with each nucleus) are displayed on
figure~\ref{fig:approx-small-Z1}.
The term labeled $(1)$ in this figure corresponds to the creation of
an on-shell electron and an off-shell positron. The positron is
subsequently put on its mass-shell by an additional interaction. For
the process $(2)$, both the electron and the positron are first
created off-shell, and then interact independently to go on-shell.

This approximation, which leads to an analytically tractable
expression, suffers however from several limitations. The obvious one
is that for the collision of two nuclei like gold for RHIC, the charge
of the nuclei is not small enough to justify this approximation, and
the inequality (\ref{eq:condZ}) is not satisfied\footnote{For the case
$Z_1\sim Z_2$, a term has to be added were the roles of $Z_1$ and
$Z_2$ are exchanged.  The accuracy of the result obtained via this
procedure is claimed to be better than the percent level for RHIC
energies \cite{IvanoSS1,IvanoSS2}.}. Moreover, this approximation
  leaves out some terms that seem physically important. In particular,
  it does not include any term where both the $e^+$ and the $e^-$ in the
  pair interact with the two nuclei. Such a term would indeed have four
  scattering matrices.

  \subsection{Limitation in the number of intermediate pairs}
  Another way to approximate the Feynman propagator is to start from the
  definition
  \begin{equation}
  G_F(x,y)\equiv {{\left<0_{\rm out}|{\rm T}\overline{\psi}(y)\psi(x)|0_{\rm in}\right>}\over{\left<0_{\rm out}|0_{\rm in}\right>}}\; .
  \end{equation}
  Then, one can use a complete\footnote{Strictly speaking, these states
    form a basis of the subspace containing only Fock states with zero
    electric charge. This is sufficient here since they will be
    contracted with the vacuum.} set of states $\{\left|n_{\rm
      in}\right>\}_{n=0\cdots\infty}$ ($n$ counts the number of pairs,
  other continuous indices like the momenta of the particles have not
  been written explicitly) and insert in the previous equation an
  identical operator constructed as
  \begin{equation}
  {\imb 1}=\sum_{n=0}^{+\infty} \left|n_{\rm in}\right>\left<n_{\rm in}\right|\; .
  \end{equation}
  Separating the contribution from $n=0$, we obtain
  \begin{eqnarray}
  G_F(x,y)=&&\left<0_{\rm in}|{\rm T}\overline{\psi}(y)\psi(x)|0_{\rm in}\right>
  \nonumber\\
  &&+\sum_{n=1}^{+\infty} 
  {{\left<0_{\rm out}|n_{\rm in}\right>}
  \over{\left<0_{\rm out}|0_{\rm in}\right>}}
  \left<n_{\rm in}|{\rm T}\overline{\psi}(y)\psi(x)|0_{\rm in}\right>\; .
  \label{eq:insert-1}
  \end{eqnarray}
  In this formula, $n$ can be seen as the number of extra pairs produced
  (and then destroyed) in the course of the evolution of the electron.
  Another way to simplify the Feynman propagator is therefore to
  truncate the previous sum, and keep only the term obtained with $n=0$,
  that is
  \begin{equation}
  G_F(x,y)\approx 
  \left<0_{\rm in}|{\rm T}\overline{\psi}(y)\psi(x)|0_{\rm in}\right>\; .
  \end{equation}

  One can note that both the exact Feynman propagator $G_F$ and the
  approximation $\left<0_{\rm in}|{\rm
      T}\overline{\psi}(y)\psi(x)|0_{\rm in}\right>$ are solutions of
  the Dirac equation $(i\slpartial_x-e\,\slA(x)-m)G(x,y)=\delta(x-y)$. Their
  difference is therefore a solution of the homogeneous Dirac equation.
  It is in fact immediate to verify that all the terms with $n\ge 1$ in
  the right hand side of Eq.~(\ref{eq:insert-1}) are solutions of the
  homogeneous Dirac equation.

  The calculation of section \ref{sec:keldysh} in fact gives the answer
  for this correlator in terms of retarded and advanced propagators
  only: we have an exact expression for the Fourier transform
  $G_{++}(q,p)$ of $\left<0_{\rm in}|{\rm
      T}\overline{\psi}(y)\psi(x)|0_{\rm in}\right>$, which reads
  \begin{equation}
  G_{++}(q,p)
  ={1\over2}\Big[G_R(q,p)+G_A(q,p)+G_S(q,p)\Big]\; ,
  \label{eq:approx-++-1}
  \end{equation}
  with $G_S(q,p)$ given in Eq.~{(\ref{eq:Gt-1})}. We see that
  contrary to the exact $G_F(q,p)$, this correlator can be written in
  closed form, in terms of the exactly known $G_R$ and $G_A$:
  \begin{eqnarray}
  &&\!\!\!\!\!\!\!\!\!\!\!\!\!G_{++}(q,p)
  ={1\over 2}\Big[
  G_R(q,p)+G_A(q,p)\nonumber\\
  &&
  +\int{{d^4k}\over{(2\pi)^4}}2\pi\delta(k^2-m^2)\;
  G_R^0(q){\cal T}_R(q,k)(\slk+m){\cal T}_A(k,p)G_A^0(p)
  \Big]\; ,
  \label{eq:approx-++-2}
  \end{eqnarray}
  where ${\cal T}_R$ and ${\cal T}_A$ are the retarded and advanced
  scattering matrices in presence of the two nuclei. Each of them can
  contain up to two of the 1-nucleus scattering matrices $T_{R,A}$ (see
  Eq.~(\ref{eq:full-ret})).

  \begin{figure}[htbp]
  \centerline{\resizebox*{5cm}{!}{\includegraphics{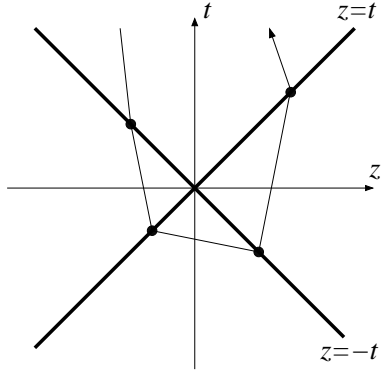}}}
  \caption{\sl New contribution to pair production in the approximation
    $G_F\approx \left<0_{\rm in}|{\rm T}\overline{\psi}(y)\psi(x)|0_{\rm
        in}\right>$. The thin lines are free propagators, the black dots
    are scattering matrices.}
  \label{fig:approx2}
  \end{figure}

  A common feature of the two expressions Eqs.~(\ref{eq:Gt-1}) and
  (\ref{eq:approx-++-1}) and Eq.~(\ref{eq:approx-++-2}) for the
  approximation of $G_F$ by $\left<0_{\rm in}|{\rm
      T}\overline{\psi}(y)\psi(x)|0_{\rm in}\right>$ is that they
  contain terms involving up to four scattering matrices on individual
  nuclei. Therefore, this approximation contains a little more than the
  previous one (based on $Z_1\alpha \ll 1$), since it includes a
  selected subset of the terms of order four in the Watson series.  The
  physical meaning of those terms is the following: an off-shell pair is
  first created. Then, the electron and the positron independently
  scatter off each of the two nuclei. No additional pair is created or
  annihilated. These terms are illustrated in figure \ref{fig:approx2}.

  \section{Conclusions}
  \label{sec:conclusions}
  In this paper, we have studied various aspects of the problem of pair
  production in the collision of ultra-relativistic heavy ions, focusing
  in resolving the discrepancies of the existent literature. 
  
  By showing that the inclusive cross-section of pair production is
  related to retarded amplitudes, while the Feynman amplitude gives
  the exclusive cross-section of single pair production, we found that
  the discrepancy between two methods used to attack the problem of
  pair production lies at a deeper level than expected: these two
  methods do not calculate the same physical quantity, and their
  result should not be compared directly.

  Then, we have studied the propagator of an electron in the
  electromagnetic background field created by two colliding nuclei. It
  appeared that the exact retarded propagator can be obtained in closed
  form if the nuclei are ultra-relativistic. However, such a
  simplification does not occur for the Feynman propagator, which can
  only be expressed as an infinite series. This problem arises in any
  background field that can produce pairs of particles, and can be
  traced back into the relativistic nature of the wave equation
  governing the electron field. This observation indicates that in the
  collision of two ultra-relativistic nuclei, the inclusive
  cross-section of pair production can be obtained exactly, but not the
  more exclusive ones. Experimentally, it would therefore be desirable
  to measure both types of cross-sections.

  Within the model of \ref{sec:model}, the inclusive cross-section,
  which is expressed in terms of the exactly known retarded amplitude,
  seems to contain Coulomb corrections as well. However, the calculation
  of completely integrated cross-sections is questionable in this
  model. Indeed, the problem raised by Lee and Milstein occurs at zero
  momentum transfer (or at infinite impact parameter), precisely where
  the model is expected to break down.  In particular, any infrared
  cutoff on the momentum transfer, coming either from an improved
  theoretical model or from experimental cuts, could drastically reduce
  these corrections.

  We have also analyzed the unitarity puzzle, and shown that pair
  production probabilities satisfy all requirements of unitarity if the
  factor $|\left<0_{\rm out}|0_{\rm in}\right>|^2$ is correctly taken
  into account. In addition, since everything depends on the 2-point
  Feynman Green's function, unitarity is preserved if one starts from an
  approximation of this propagator. A side product of this analysis is
  that the multiplicity distribution is not Poissonian.

  Finally, in addition to the approximation $Z_1\alpha\ll 1\sim Z_2$
  used in \cite{IvanoSS1,IvanoSS2}, we have presented a completely
  different approximation for $G_F(x,y)$ that leads also to closed
  expressions, and seems to contain more of the relevant physics for
  pair production by two heavy nuclei.

  \vskip 2mm
  \noindent{\bf Acknowledgements:} This work is supported by DOE under
  grant DE-AC02-98CH10886.  A.P. is supported in part by the
  A.-v.-Humboldt foundation (Feo\-dor-Lynen program). We would also like
  to thank R. Venugopalan and C. Bertulani for interesting
  discussions. We thank U.~Eichmann for pointing out the fact that
  Coulomb corrections may be present in the inclusive cross-section as
  well.

  \bibliographystyle{unsrt}

\end{document}